%% file: main.tex
\documentclass{article}

\usepackage{PRIMEarxiv}

\usepackage[utf8]{inputenc} 
\usepackage[T1]{fontenc}    
\usepackage{hyperref}       
\usepackage{url}            
\usepackage{booktabs}       
\usepackage{amsfonts}       
\usepackage{nicefrac}       
\usepackage{microtype}      
\usepackage{lipsum}
\usepackage{fancyhdr}       
\usepackage{graphicx}       
\graphicspath{{media/}}     
\usepackage{graphicx}  
\usepackage{epstopdf}
\usepackage{ragged2e}
\usepackage{algorithm}
\usepackage{algpseudocode}
\usepackage{enumerate}
\usepackage{subfigure}
\usepackage{amsmath}
\usepackage{bm}
\usepackage{stfloats}
\usepackage{siunitx}
\usepackage{booktabs}
\usepackage{multirow}
\usepackage{pifont}
\usepackage[justification=centering]{caption}

\title{Regional-Local Adversarially Learned One-Class Classifier Anomalous Sound Detection in Global Long-Term Space}

\author{
  Yu Sha \\
   Xidian University\\
   FIAS \thanks{\textit{FIAS: Frankfurt Institute for Advanced Studies}}\\
   XF-IJRC\thanks{\textit{XF-IJRC: Xidian-FIAS international Joint Research Center}}\\
   \And
  Johannes Faber \\
  FIAS \\
    \And
  Shuiping Gou \\
  Xidian University \\
    \And
  Bo Liu \\
  Xidian University \\
    \And
  Wei Li \\
  FIAS \\
    \And
  Stefan Schramm \\
  FIAS \\ 
    \And
  Horst Stoecker \\
  FIAS \\  
  Goethe Universit{\"a}t \\
  GSI\thanks{\textit{GSI: GSI Helmholtzzentrum f{\"u}r Schwerionenforschung GmbH}}\\
    \And
  Thomas Steckenreiter \\
  SAMSOM AG \\  
    \And
  Domagoj Vnucec \\
  SAMSOM AG \\  
    \And
  Nadine Wetzstein \\
  SAMSOM AG \\      
  \And
  Andreas Widl \\
  SAMSOM AG \\  
  \And
  Kai Zhou \thanks{\textit{Kai Zhou is the corresponding author. Email: zhou@fias.uni-frankfurt.de}}\\
  FIAS \\  
}

\begin{document}
\maketitle

\input{sec/0_abstract}


\input{sec/1_introduction}

\input{sec/2_preliminaries}
\input{sec/3_method}

\input{sec/4_experiments}

\input{sec/5_relatedwork}
\input{sec/6_conclusions}

\input{sec/Acknowledgements}

\bibliographystyle{unsrt}
\bibliography{main}
\clearpage
\appendix
\input{sec/supplementary}

\end{document}

%% file: sec/0_abstract.tex
\begin{abstract}
Anomalous sound detection (ASD) is one of the most significant tasks of mechanical equipment monitoring and maintaining in complex industrial systems. In practice, it is vital to precisely identify abnormal status of the working mechanical system, which can further facilitate the failure troubleshooting. In this paper, we propose a multi-pattern adversarial learning one-class classification framework, which allows us to use both the generator and the discriminator of an adversarial model for efficient ASD. The core idea is learning to reconstruct the normal patterns of acoustic data through two different patterns of auto-encoding generators, which succeeds in extending the fundamental role of a discriminator from identifying real and fake data to distinguishing between regional and local pattern reconstructions. Furthermore, we present a global filter layer for long-term interactions in the frequency domain space, which directly learns from the original data without introducing any human priors. Extensive experiments performed on four real-world datasets from different industrial domains (three cavitation datasets provided by SAMSON AG, and one existing publicly) for anomaly detection show superior results, and outperform recent state-of-the-art ASD methods.
\end{abstract}

%% file: sec/1_introduction.tex
\section{Introduction}
\label{sec:intro}
Anomaly detection has been extensively studied in different domains \cite{chalapathy2019deep} (e.g., image, video and time series, etc.). Due to the rare occurrence of anomalous scenarios, the anomaly detection problem is often called one-class classification (OCC) in which only normal data is present and used to learn a novelty detection model. In this paper, we focus on anomalous sound detection (ASD) \cite{koizumi2018unsupervised} for cavitation or other faults in complex industrial mechanical systems, which has been an active research topic in industry during these years. ASD is the task of determining whether the acoustic signal emitted from a target machine is normal or abnormal, which is a fundamental technology in the fourth industrial revolution to monitor the health of machines by `listening' to their acoustics \cite{koizumi2020description}.

Conventionally, for anomaly detection in industry, engineering experts manually establish static thresholds for abnormal status based on physical definitions of cavitation \cite{franc2006fundamentals} or mechanical structure of the target machine \cite{wei2019review}. However, this process requires huge human and economic costs as the scale and the complexity of industrial data grows significantly over the years. To tackle this problem, many unsupervised anomaly detection algorithms have been developed for vibration signals or time series \cite{nguyen2021deep}, where anomalies are detected mainly for one specific state. However, for a complex real-world industrial system, the anomaly states are often complicated, diversified and interacted with each other (e.g., multiple states of cavitation, leakage, contamination, etc.) due to their different mechanical design configurations and different sensors. Therefore, specific single-state anomaly detection is no longer compatible with the current complex real-world industrial system.

Formally, ASD data consist of multiple measured signal events, each of which describes an anomalous state of a complex entity and each records the entire physical process from the beginning to the end of an anomalous state. Thus, there is global long-term dependency of each entity measurement event. Violating the dependency of each entity measurement event would lead to degradation of anomaly detection performance. To help the system operates better in anomalies monitoring for complex industrial mechanical equipment, the algorithm should consider the global long-term dependency of measurement events \cite{bengio1993problem}.

Recent methods for ASD can be roughly divided into two classes: encoder-decoder based reconstruction and adversarial training based reconstruction. Encoder-decoder-based methods \cite{baldi2012autoencoders} try to train a model to produce good quality reconstructions of input signals and detect abnormal data based on reconstruction loss. For adversarial training-based approachs \cite{tian2021analysis}, it can substantially improve the quality of reconstruction. At test time, the trained generator $\mathcal{G}$ is then uncoupled from the discriminator $\mathcal{D}$ to be used as a reconstruction model. As mentioned in \cite{sabokrou2018adversarially}, the reconstruction loss between normal and abnormal data is significantly different due to the adversarial training, which gives rise to good anomaly detection results. However, in some cases currently using only the reconstruction capability of a generator $\mathcal{G}$ is useless because of the spillover of the reconstruction capability of the generator $\mathcal{G}$, which thus largely degrades the performance of anomaly detection. 
\input{fig_input/input_differentpatterns}

Based on the above considerations, for industrial acoustic signals anomaly detection, we propose to explicitly learn global long-term dependencies and multi-pattern high quality reconstructions to better capture the normal patterns of industrial acoustic signals. To address the global long-term dependence problem, a global filter layer is introduced and applied to the spectrum of the input features. And, the global filter layer is directly learned from the original sub-sequence acoustic signals without introducing extra human priors. Since the global filter layer is able to cover all frequencies, our method therefore capture the global long-term interactions. To address multi-pattern high quality reconstructions, we devise an adversarial learning approach with multiple generators and a single discriminator (see Figure \ref{fig: DifferentPatterns}). The two generators $\mathcal{G} ^{local}$ and $\mathcal{G} ^{regional}$ try to generate multi-pattern real-looking fake data, and they focus on the local and regional details of acoustic signals under the global long-term space, respectively. Furthermore, the capability of discriminator $\mathcal{D}$ is extended from distinguishing between real and fake to differentiating between local and regional reconstructions, making it more suitable for anomaly detection task on industrial acoustic signals. In addition, we break the conventional idea of using only $\mathcal{G}$ or $\mathcal{D}$ for anomaly detection. In our method, the $\mathcal{G} ^{local}+\mathcal{G} ^{regional}+\mathcal{D}$ are used simultaneously for anomaly detection. Finally, the above form our data-driven adversarial training anomaly detection approach, namely \underline{R}egional-\underline{L}ocal Adversarial Learning One-Class Classifier Anomalous Sound Detection in \underline{G}lobal Long-Term Space (GRLNet) for cavitation detection of acoustic signals in complex industrial systems. 

The contributions of this paper are summarised as follows:
\begin{itemize}
    \item To the best of our knowledge, our proposed GRLNet is the first adversarial learning one-class classifier for industrial acoustic signals anomaly event detection which employs $\mathcal{D}$ along with $\mathcal{G}$.
    \item We propose a global filter layer based on a global filter (1D FFT) for long-term interactions in the frequency domain space, which directly learns from the original data without introducing extra human priors.
    \item We extend the capability of $\mathcal{D}$ from discriminating real signals against fake ones to differentiating between local and regional reconstructions by producing different patterns of reconstructed signals using $\mathcal{G} ^{local}$ and $\mathcal{G} ^{regional}$.
    \item Our proposed GRLNet outperforms state-of-the-art methods in the experiments conducted on four real-world acoustic datasets from different industrial domain (our own three cavitation datasets provided by SAMSON and one public dataset). Moreover, ablation studies further demonstrate the effectiveness of our proposed structure design for industrial acoustic anomaly detection. We publish our code for better reproducibility.
\end{itemize}

%% file: fig_input/input_differentpatterns.tex
\begin{figure}
    \centering
    \includegraphics[width=0.45\textwidth,height=45mm]{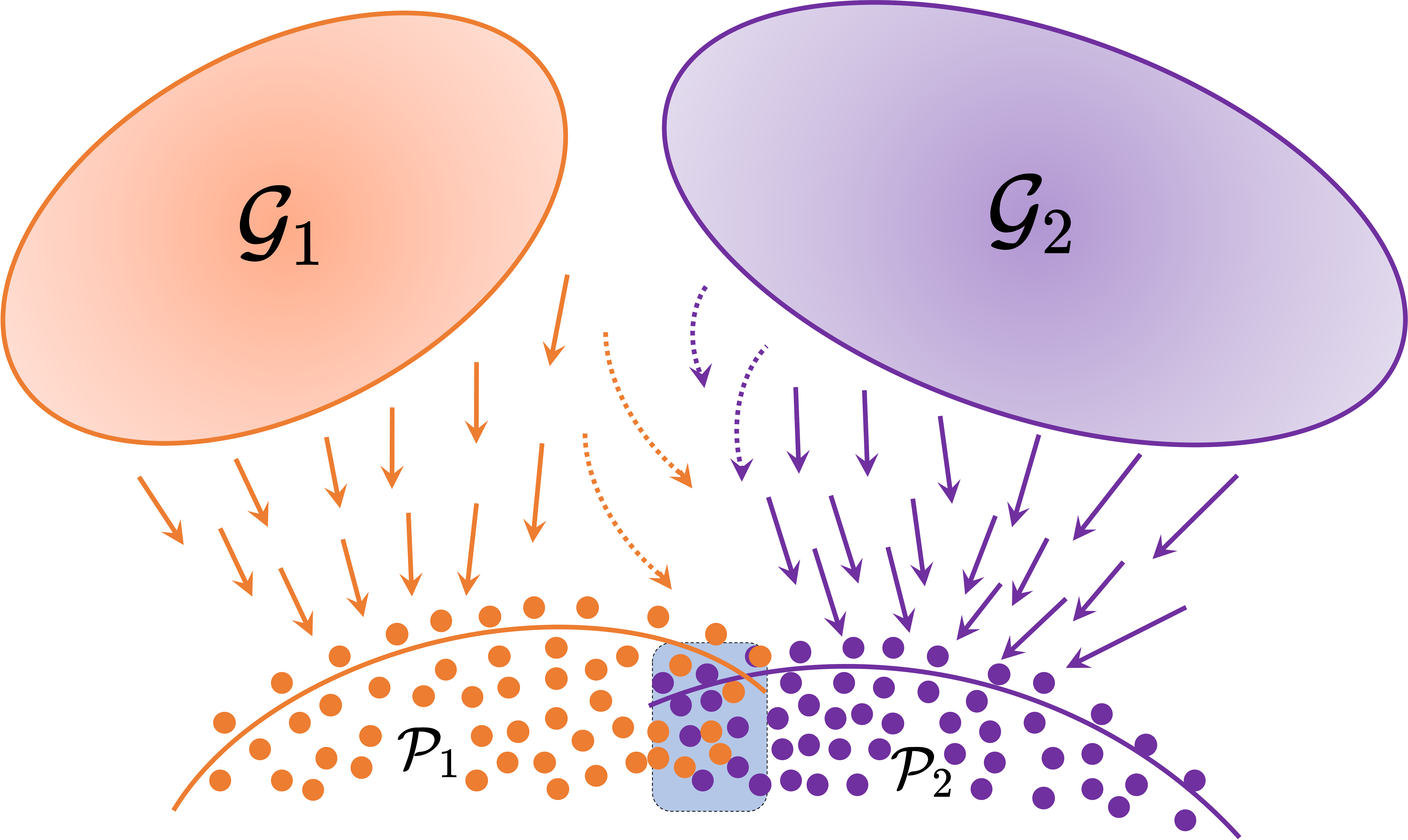}
    \caption{Visualisation of different generators produce different patterns. $\mathcal{G}_{1}$ and $\mathcal{G}_{2}$ represent different generators. $\mathcal{P}_{1}$ and $\mathcal{P}_{1}$ are two clusters with different patterns and the blue area indicates the possible overlap of the two clusters. The arrows abstractly denote generator gradients.}
    \label{fig: DifferentPatterns}
\end{figure}

%% file: sec/2_preliminaries.tex
\section{PRELIMINARES}
\label{sec: preliminaries}

\subsection{Abnormal Cavitation Event Detection}
\label{sec: Abnormal Cavitation Event Detection}
Cavitation is defined as the phenomenon that bubbles form on the solid surface when the pressure at the contact point of the liquid and the solid surface is lower than its vapor pressure \cite{jazi2009detecting,bonnier2002experimental}. For working valves, the cavitation acoustic signals of four different flow states are obtained as continuous waveforms using acoustic sensors as shown in Figure \ref{fig: different cavitations}. Each of the observed acoustic signals record the whole physical process from the beginning to the end of the event of corresponding cavitation state. For complex industrial systems, cavitation can be roughly classified into three types: cavitation choked flow, constant cavitation and incipient cavitation (see (a)-(c) in Figure \ref{fig: different cavitations}). For incipient cavitation, only a few vapor bubbles are generated inside pipes or valves and bubbles implode with noise, which indicates the beginning of an abnormal system. In constant cavitation, more vapor bubbles are produced from interior of pipes or valves and accompanied by a certain level of noise, which means the system is completely abnormal. When both vapor bubbles concentration and noise inside pipes or valves reach maximum value which leads to a choked flow condition with cavitation. At this situation, the system is almost impossible to operate and most of abnormalities are actually easier to be detected. No matter severe or subtle, each type of cavitation can indicate a potential problem in the system. Therefore, precisely detecting such abnormal cavitation events is urgently requested for operators of industrial systems.
\input{fig_input/input_differentcavitation}

\subsection{Acoustic Signals Augmentation}
\label{sec: data augmentation}
Formally, we suppose there is $x\subseteq {\mathcal{R}}^{M\times N}$ with $M$ measurements for each acoustic signal. Considering purposely steady flow status (i.e. it’s always the same fluid status class within the individual measurement duration with 3 $sec$ or 25 $sec$ time-length) in each recorded stream and fine resolution for the sensor. One can actually split each stream into several pieces which still can hold enough essential information for detection. And, with independent features per piece due to the intrinsic randomness of the noise emission, given the piece is not so short. Therefore, according to the above theory, we apply a sliding window with window size ${s}_{w}$ and step size ${s}_{s}$ to divide the acoustic signal sequence into a set of sub-sequences $X=\left\{\bm{{x}}_{i,j},i=1,2,\ldots,N;j=1,2,\ldots,k\right \}\subseteq {\mathcal{R}}^{{s}_{w}\times N}$, where $k=\frac{\left(M-{s}_{w}\right)}{{s}_{s}}$ is the number of sub-sequences (as shown in the top left of Figure \ref{fig: framework}) and $N$ is stream. 

\subsection{Discrete Fourier Transform}
\label{sec:1D FFT}
The Discrete Fourier Transform (DFT) \cite{madisetti1997digital} converts a signal from time domain space to frequency domain space, which is an essential part of the global filtering layer in GRLNet. Given a sequence of signal $\left \{{x}_{n}\right\}_{n=0}^{N-1}$, the 1D DFT converts the signal sequence into the frequency domain is defined by the formula:
\begin{equation}
\label{eq: 1D DFT}
X[k] = \sum\limits_{n = 0}^{N - 1} {{x_n}{e^{ - \frac{{2\pi j}}{N}nk}}}  := \sum\limits_{n = 0}^{N - 1} {{x_n}W_N^{kn}} 
\end{equation}
where $j$ is the imaginary unit and ${W_N} = {e^{ - \frac{{2\pi j}}{N}}}$. The formulation of 1D DFT of Equation \ref{eq: 1D DFT} can be derived from the Fourier transform for continuous signals (see Appendix \ref{sec: Discrete Fourier transform} for details). The 1D DFT generates a sequence $X[k]$ as the sum of all the original input signals ${x}_{n}$ for each $k$. And $X[k]$ represents the spectrum value of the original signal sequence ${x}_{n}$ at the discrete frequency point ${w_k} = {{2\pi k} \mathord{\left/{\vphantom {{2\pi k} N}} \right.\kern-\nulldelimiterspace} N}$. Specifically, $X[k]$ is equivalent to applying an equal interval sampling to the spectrum $X[{e^{jw}}]$ in the range $\left[ {0,2\pi } \right)$ with a sampling interval $\Delta w = {{2\pi } \mathord{\left/{\vphantom {{2\pi } N}} \right.\kern-\nulldelimiterspace} N}$. 

There are two main reasons why DFT is widely applied to signal processing: (1) physically the spectrum in frequency domain space reveals more clearly the acoustic dynamics than the original waveforms; (2) the inputs and outputs of DFT are discrete and can be easily computed by computers, and there exist efficient ways for the computation of the DFT. There are two main approaches to compute the DFT: (1) the Fast Fourier Transform (FFT); (2) the matrix multiplication method. Cooley-Tukey \cite{johnsson1992cooley} is the standard FFT algorithm, which is divided into a time extraction method and a frequency extraction method. Due to the symmetry and periodicity of the DFT, the computational complexity of the DFT is reduced from $\mathcal{O}({N}^{2})$ to $\mathcal{O}(N\log N)$. 

%% file: fig_input/input_differentcavitation.tex
\begin{figure}[htbp]
\centering
\subfigure[Cavitation Choked Flow]{
\begin{minipage}[t]{0.45\linewidth}
\centering
\includegraphics[width=\textwidth,height=35mm]{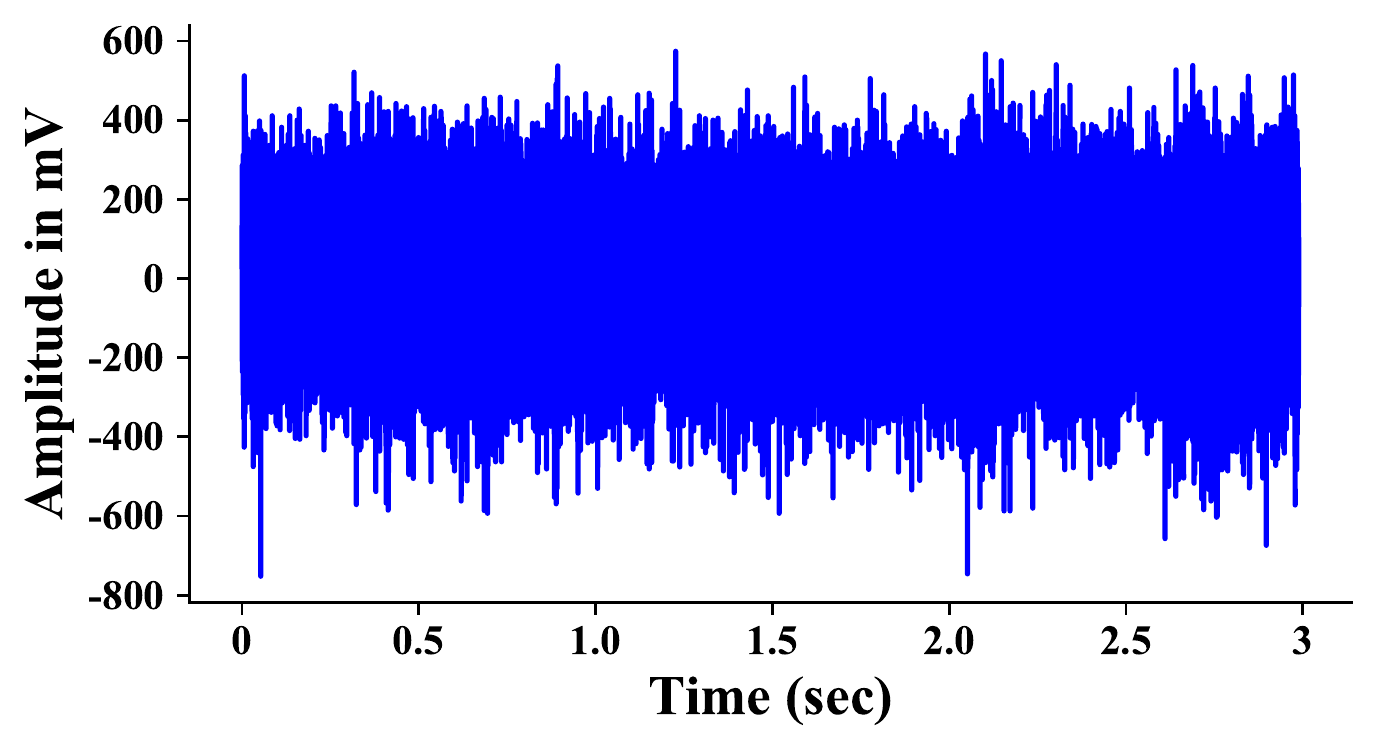}
\end{minipage}%
}%
\subfigure[Constant Cavitation]{
\begin{minipage}[t]{0.45\linewidth}
\centering
\includegraphics[width=\textwidth,height=35mm]{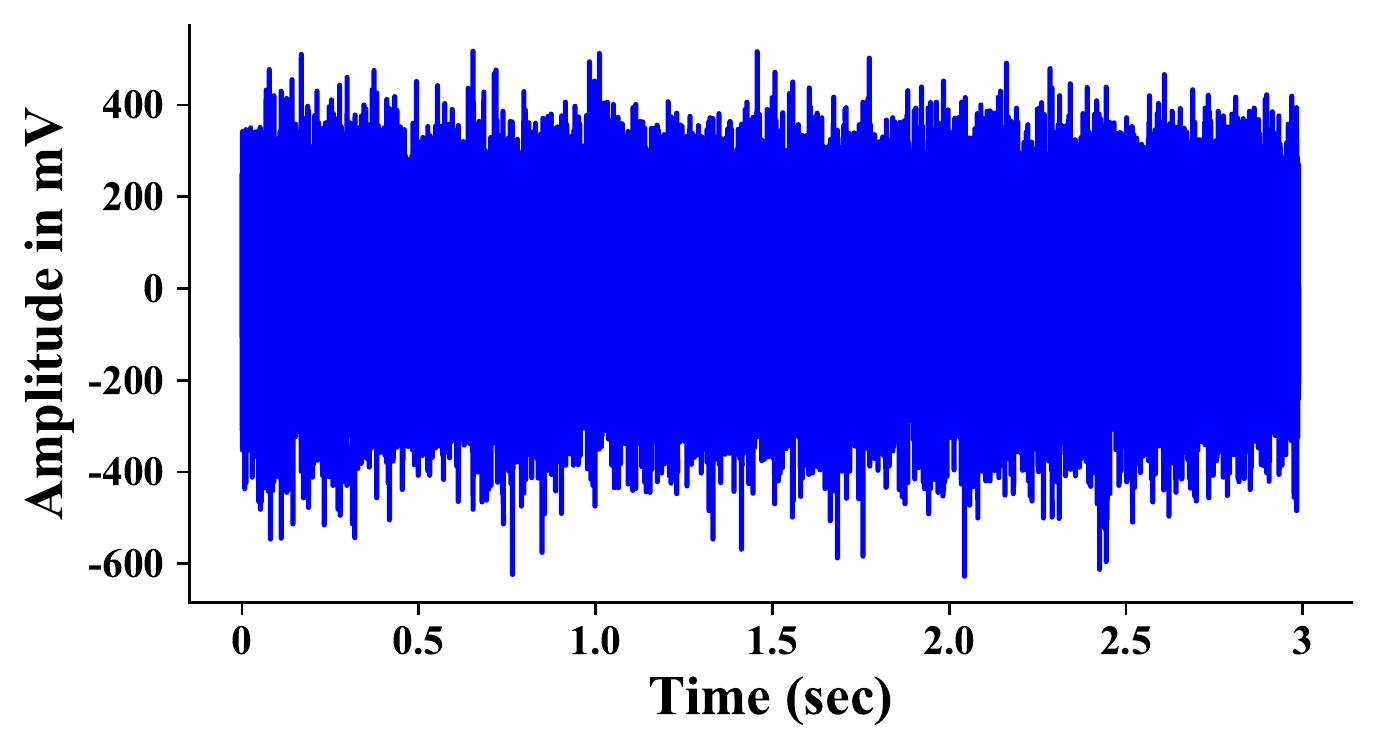}
\end{minipage}%
}%

\subfigure[Incipient Cavitation]{
\begin{minipage}[t]{0.45\linewidth}
\centering
\includegraphics[width=\textwidth,height=35mm]{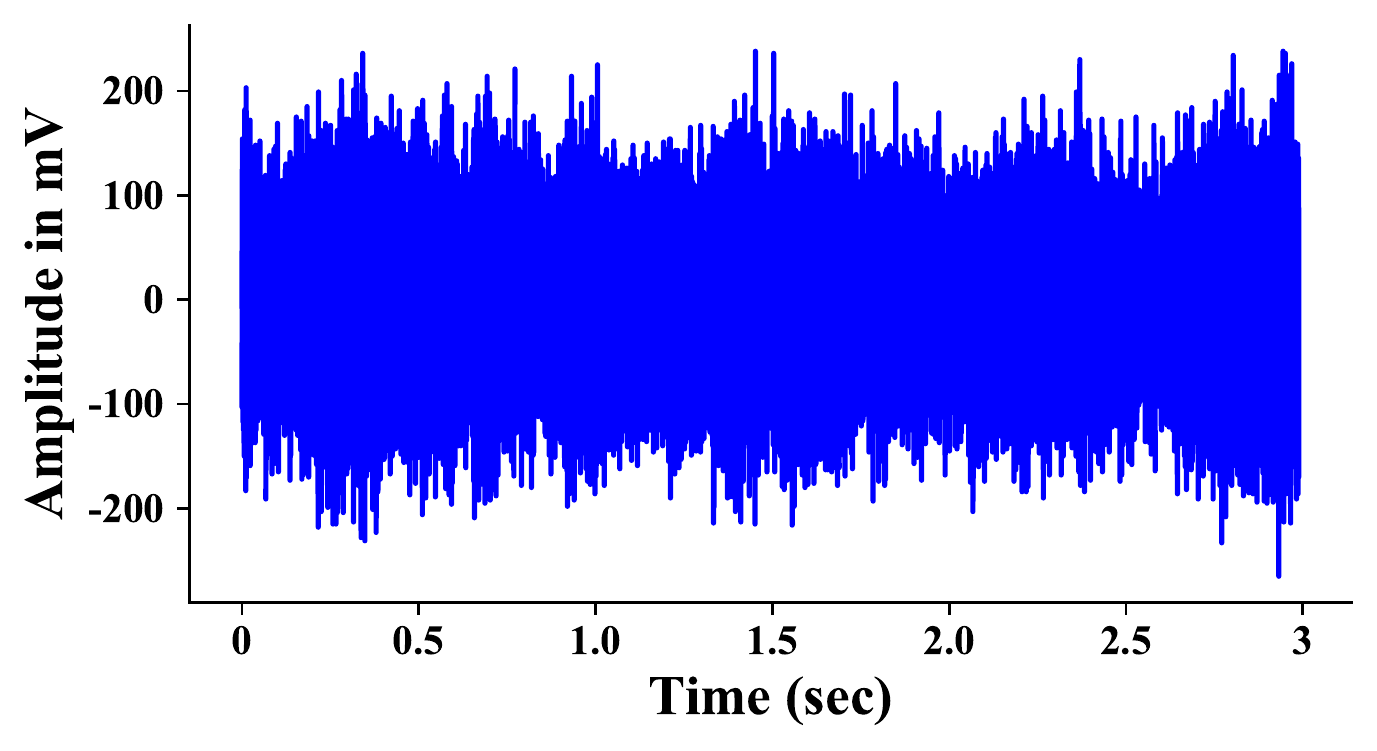}
\end{minipage}%
}%
\subfigure[Non cavitation]{
\begin{minipage}[t]{0.45\linewidth}
\centering
\includegraphics[width=\textwidth,height=35mm]{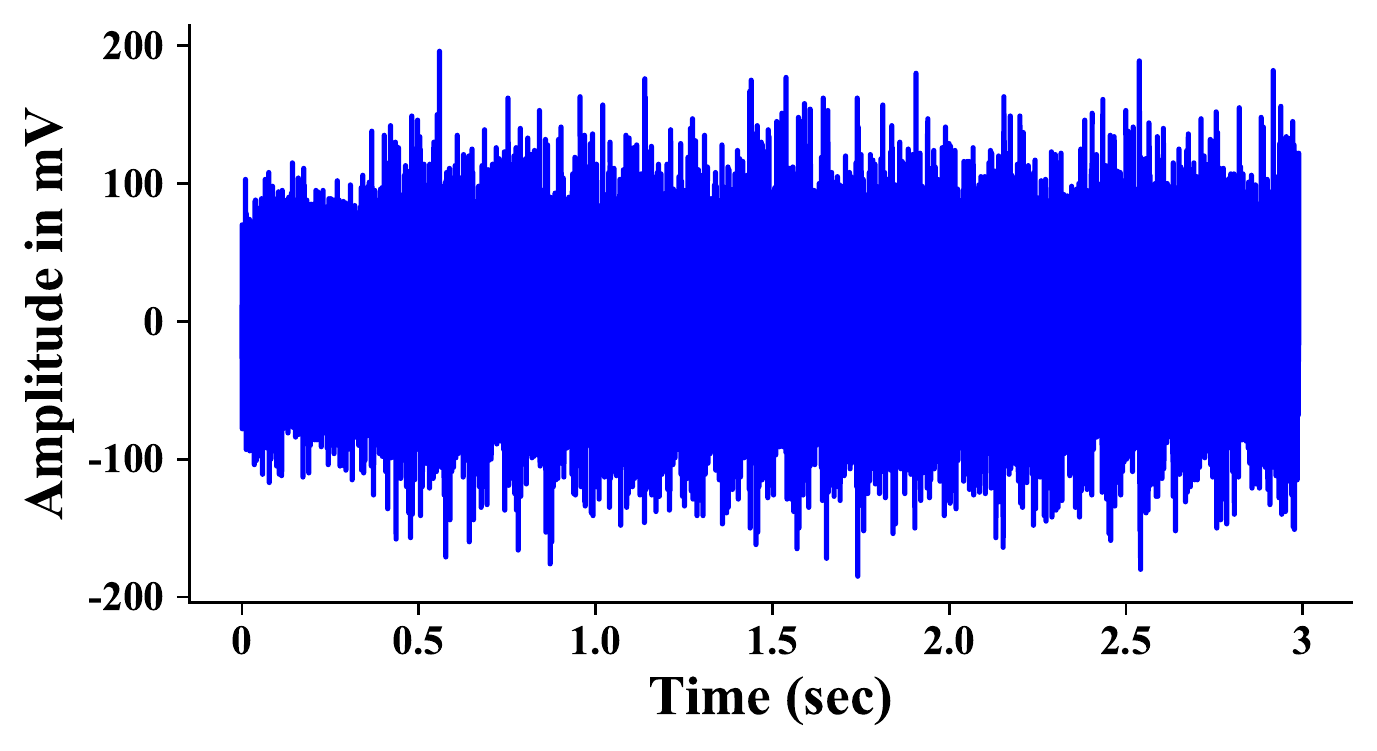}
\end{minipage}%
}%
\centering
\caption{Acoustic signals in (a)-(d) for cavitation choked flow, constant cavitation, incipient cavitation and non-cavitation, respectively.}
\label{fig: different cavitations}
\end{figure}

%% file: sec/3_method.tex
\section{Method}
\begin{figure}
    \includegraphics[width=0.95\textwidth,height=75mm]{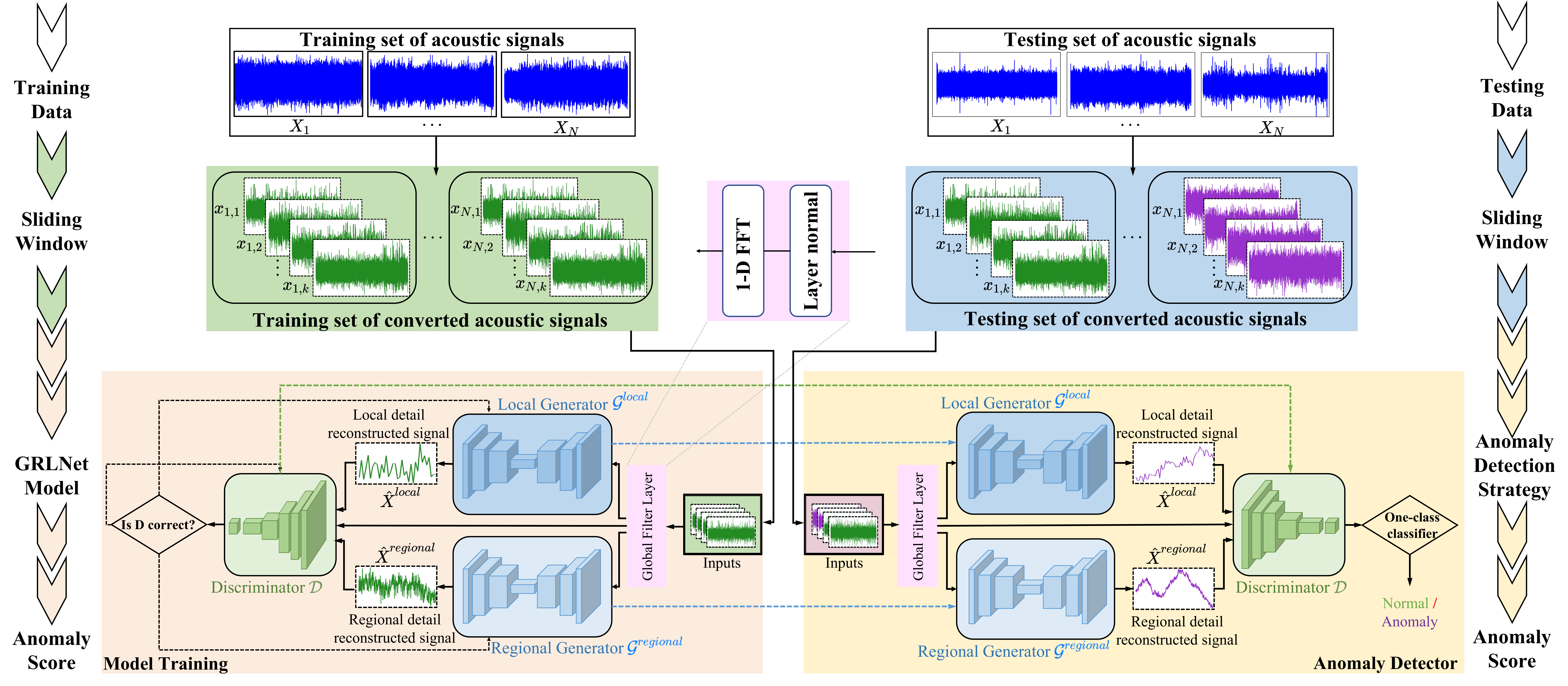}
    \centering
    \caption{Our proposed GRLNet framework. In the training phase, the segmented acoustics signals are fed through the global filter layer (involving layer normalization and 1D FFT) to $\mathcal{G} ^{regional}$, $\mathcal{G} ^{local}$ and $\mathcal{D}$ to obtain a reasonably trained state of $\mathcal{G} ^{regional}$, $\mathcal{G} ^{local}$ and $\mathcal{D}$. Good quality examples correspond to real training of segmented acoustics signals, local reconstructed acoustic signals acquired by $\mathcal{G} ^{local}$ and regional reconstructed acoustic signals obtained from $\mathcal{G} ^{regional}$. During testing, the to be detected segmented acoustics signals are inferred through $\mathcal{G} ^{regional}$, $\mathcal{G} ^{local}$ and $\mathcal{D}$ after passing through the global filter layer and the output of $\mathcal{D}$ is considered as anomaly score.}
    \label{fig: framework}
\end{figure}

\subsection{Architecture Overview}
The structure of the generator and discriminator in GRLNet is similar to those proposed by Sabokrou et al \cite{sabokrou2018adversarially}. The generator is a 1D convolutional autoencoder, which is coupled with a discriminator to learn one class of data in an unsupervised adversarial manner. GRLNet has two generators, $\mathcal{G} ^{local}$ and $\mathcal{G} ^{regional}$, and a discriminator $\mathcal{D}$ (more details see Appendix \ref{sec: AppendixExperiments}). The $\mathcal{G} ^{local}$ and $\mathcal{G} ^{regional}$ focus on local and regional reconstruction of the signal, respectively, which can capture different patterns of the real signal distribution at the same time. The training of the model is using min-max strategy to optimize the following overall objective function.
\begin{equation}
\label{eq: objective function}
\begin{array}{l}
\mathop {\min }\limits_{{\theta _g}} \mathop {\max }\limits_{{\theta _d}} V({\theta _d},{\theta _g}): = {\mathbb{E}_{X \sim {p_t}}}[\log (1 - \mathcal{D}(X;{\theta _d}))]\\\,\,\,\,\,\,\,\,\,\,\,\,\,\,\,\,\,\,\,\,\,\,\,+{\mathbb{E}_{\tilde{X} \sim {p_t} + {N_\sigma }}}[\log (\mathcal{D}(\mathcal{G}(\tilde X;{\theta _g});{\theta _d}))]
\end{array}
\end{equation}
where $X$ denotes a set of true samples with distribution ${p_t}$; ${\tilde X}$ is the input sample $X$ with added noise ${\mathcal{N}}_{\sigma}$; ${\theta _g} = ({\theta _{{g^{local}}}},{\theta _{{g^{regional}}}})$ and ${\theta _d}$ represent learnable parameters of the $\mathcal{G}$ and $\mathcal{D}$, respectively. The above objective function is optimised in a block-wise manner, i.e. one of ${\theta _g}$ and ${\theta _d}$ is optimised while the other is fixed. Theoretically, the generator could produce real samples in Nash-equilibrium, which means ${p_t}={p_g}$ (${p_g}$ is distribution from the generator).

The overall architecture of our model is depicted in Figure \ref{fig: framework}. In our model, we take the non-overlapping acoustics signal patches of length $L$ obtained through the sliding window as input. The basic building blocks of GRLNet consist of: (1) a global filter layer of complexity $\mathcal{O}(L\log L)$ that captures global features of acoustics signals; (2) a local generator $\mathcal{G} ^{local}$ and a regional generator $\mathcal{G} ^{regional}$ that can simultaneously capture different patterns of the reconstructed acoustics signals; (3) a 1D convolution discriminator $\mathcal{D}$. In this way, the final purpose of our proposed GRLNet anomaly detection framework is to extend the capabilities of discriminator $\mathcal{D}$ from distinguishing between real and fake to differentiating between local and regional reconstructions. 

\subsection{Training}
Given a training dataset ${\mathcal{X}}^{train}\subseteq{\mathcal{R}}^{M\times N}$ and a test dataset ${\mathcal{X}}^{test}\subseteq{\mathcal{R}}^{M\times N}$ with $N$ streams and $M$ measurements for each stream. Our task is to assign binary labels to the acoustic signals of test dataset. Note that the acoustic signals in the training dataset are normal.

In the training stage, the proposed GRLNet is similar to common practices in training adversarial one-class classifiers \cite{wang2021multi,liu2012isolation}. $X$ is passed through a global filter layer to obtain global frequency features of acoustics signals. Then, $\mathcal{G} ^{local}$ attempts to regenerate real-looking fake data which focuses on the local details of acoustics signals and $\mathcal{G} ^{regional}$ tries to regenerate real-looking fake data which focuses on the regional details of acoustics signals, which are then fed into $\mathcal{D}$ along with real data. Finally, $\mathcal{D}$ learns to discriminate between real and fake data, and also learns to reconstruct acoustics signals with different patterns generated by $\mathcal{G} ^{local}$ and $\mathcal{G} ^{regional}$. In general, this stage minimises the following loss function:
\begin{equation}
\mathcal{L}=\lambda\left({\mathcal{L}}_{\mathcal{G}+\mathcal{D}}\right)+\left(1-\lambda \right){\mathcal{L}}_{{R}^{total}}
\end{equation}
where ${\mathcal{L}}_{{\mathcal{G}}+{\mathcal{D}}}={\mathcal{L}}_{{\mathcal{G}}^{local}+\mathcal{D}}+{\mathcal{L}}_{{\mathcal{G}}^{regional}+\mathcal{D}}$ is the loss function of our joint training, and the overall objective defined in Equation \ref{eq: objective function}; ${\mathcal{L}}_{{R}^{total}}={\mathcal{L}}_{{R}^{local}}+{\mathcal{L}}_{{R}^{regional}}$ is the total reconstruction loss with ${\mathcal{L}}_{{R}^{local}}=\parallel X-G(\tilde{X}){\parallel }^{2}$ is the local reconstruction loss and ${\mathcal{L}}_{{R}^{regional}}=\parallel X-G(\tilde{X}){\parallel }^{2}$ is the regional reconstruction loss; $\lambda$ is a weighing hyperparameter, which is usually set 0.5.

\noindent\textbf{Global Filter Layer.} We propose the global filter layer to extract global frequency features, which can mix sub-sequences representing different spatial locations. Given the 
sub-sequences $\bm{x}\in X$, we first apply layer normalization \cite{ba2016layer} to normalize $\bm{x}$:
\begin{equation}
\bm{\hat{x}}=\mathcal{LN}\left[\bm{x}\right]
\end{equation}
where $\mathcal{LN}\left[\cdot\right]$ denotes the layer normalization. We can then perform 1D FFT (see Section \ref{sec:1D FFT}) to convert $\bm{\hat{x}}$ to the frequency domain:
\begin{equation}
\bm{X}=\mathcal{F}\left [\bm{\hat{x}}\right]
\end{equation}
where $\mathcal{F}\left[\cdot\right]$ represents the 1D FFT. It is to be noted that $\bm{X}$ is a complex tensor and is denoted as the spectrum of $\bm{\hat{x}}$. 

The concept of the global filter layer is motivated by frequency filters in digital signal processing \cite{madisetti1997digital}. Based on the convolutional theorem of the Fourier transform, it can be proved that the global filter layer is equivalent to the depth global circular convolution (see Appendix \ref{sec: Discrete Fourier transform}). Therefore, the global filter layer is different from the standard local convolutional layer and it can extract global features of acoustic signals.

It is worth noting that in the specific implementation, we can keep only half of the values of $\bm{X}$ to reduce redundant computation based on the conjugate symmetry of the DFT, i.e. $\bm{X}\left [L-u,:\right]=\bm{{X}}^{*}\left [L,:\right]$ (see Appendix \ref{sec: Discrete Fourier transform}). Given a real tensor $\bm{\hat{x}}$, $\bm{{X}_{h}}$ has the half of the values in the $\bm{X}$:
\begin{equation}
\bm{{X}_{h}}=\bm{X}[0:\hat{L}]:={\mathcal{F}}_{h}\left[\bm{\hat{x}}\right],\hat{L}=\left \lceil L/2\right\rceil
\end{equation}
where ${\mathcal{F}}_{h}$ denotes the 1D FFT of $\bm{{X}_{h}}$ which takes only half the values of $\bm{X}$ and retains the complete information about $\bm{X}$.

\noindent\textbf{Local and Regional Generator.} Both $\mathcal{G} ^{local}$ and $\mathcal{G} ^{regional}$ are structures for an autoencoder consisting of 1D convolution, but they generate different patterns of reconstructed signals. $\mathcal{G} ^{local}$ generates reconstructed signals focusing on local detail with smaller convolutional kernels. $\mathcal{G} ^{regional}$ generates reconstructed signals with regional detail using larger convolutional kernels.

\noindent\textbf{Goal.} The essence of our training is to provide examples of different patterns reconstructions to $\mathcal{D}$, with an aim to make it learn the kind of output that $\mathcal{G} ^{local}$ and $\mathcal{G} ^{regional}$ would produce in the case of anomalous inputs. Furthermore, the capability of $\mathcal{D}$ is extended from distinguishing between true and false to differentiating between local and regional reconstructions.

\noindent\textbf{Tweaking the objective function.} Based on the related theory of \cite{ghosh2018multi}, $\mathcal{D}$ must learn to push different generators to different recognisable patterns. And the target for each generator remains the same as in the standard GAN \cite{goodfellow2014generative}. Therefore, for the $j$-th generator ($j={1,2}$), the objective is to minimize the following.
\begin{equation}
\begin{array}{l}
{\mathbb{E}_{X \sim {p_t}}}\log (1 - {\mathcal{D}}_{k+1}(X;{\theta _d}))\\\,\,+{\mathbb{E}_{\tilde{X} \sim {p_t} + {N_\sigma }}}\log ({\mathcal{D}}_{k+1}({\mathcal{G}}_{j}(\tilde X;{\theta _g^j});{\theta _d}))
\end{array}
\end{equation}
where, at $k=1$, which has one generator. The gradient of each generator can be simply denoted as ${\nabla _{\theta _g^j}}\log ({\mathcal{D}_{k + 1}}({\mathcal{G}_j}(\tilde X;\theta _g^j);{\theta _d}))$ and all generators ($\mathcal{G} ^{local}$ and $\mathcal{G} ^{regional}$) can be updated in parallel. The gradient of $\mathcal{D}$ is ${\nabla _{{\theta _d}}}\log (1 - {\mathcal{D}_i}(X;{\theta _d}))$, where ${\mathcal{D}_i}(X;{\theta _d})$ is the $i$-th index of $\mathcal{D}(X;{\theta }_{d})$ and $i\in \left \{1,\ldots ,k\right \}$.

\subsection{Testing} In the test stage, as shown on the right in Figure \ref{fig: framework}, $\mathcal{G} ^{local}$, $\mathcal{G} ^{regional}$ and $\mathcal{D}$ are utilised for one-class classification (OCC) \cite{wang2021multi,liu2012isolation}. The final classification decision for an input acoustic signal is given as follows:
\begin{equation}
OCC=\begin{cases}
 Normal \,signal\,\,\,\,if\,\beta\mathcal{D}(\mathop{{\mathcal{G}}}\nolimits_{{local}}^{{regional}}\left(X\right))+(1-\beta)\mathcal{D}(X)<\tau \\ 
 Anomaly \,signal \,\,\,\, otherwise
\end{cases}
\end{equation}
where $\beta$ is comprehensive output factor to weight the output of $\mathcal{D}$ under different patterns and $\tau $ is a predefined threshold. The proposed GRLNet is summarized in Algorithm \ref{algo: GRLNet anomaly detection}.
\input{Algo/algorithm}

%% file: Algo/algorithm.tex
\begin{algorithm}[htbp]
	\caption{GRLNet-based Anomaly Detection Strategy}
	\label{algo: GRLNet anomaly detection}
	\begin{algorithmic}
    \Require input original sub-sequences $X\subseteq {R}^{{s}_{w}\times N}$, some weighting factors $\alpha$, $\beta$ and $\lambda$ 
    \Ensure anomaly scores by one-class classifier
    \State fake $\leftarrow$ a tensor of all elements of one with the same size as $X$
    \State real $\leftarrow$ a tensor of all elements of zero with the same size as $X$
    \For{epoch $= 0,1,\ldots,N$}
    \State Generate samples from $X$ with added noise ${\mathcal{N}}_{\sigma}$:
    \State $\tilde{X}=\left \{{\tilde{x}}_{i},i=1,\ldots ,m\right\}\Rightarrow {\mathcal{G}}^{local}(\tilde{X}),{\mathcal{G}}^{regional}(\tilde{X})$
    \State Train $\mathcal{D}$:
    \State $\mathcal{D}(X)$, $\mathcal{D}(\mathcal{G}^{local}(\tilde{X}))$, $\mathcal{D}(\mathcal{G}^{regional}(\tilde{X}))$
    \State Update parameters of $\mathcal{D}$ by minimizing ${\mathcal{D}}_{loss}$:
    \State $BCE(\mathcal{D}(\mathcal{G}^{local}(\tilde{X})),fake)\Rightarrow BCE_{fake}^{local}$
    \State $BCE(\mathcal{D}(\mathcal{G}^{regional}(\tilde{X})),fake)\Rightarrow BCE_{fake}^{regional}$
    \State $BCE(\mathcal{D}(X),real)\Rightarrow {BCE}_{real}$
    \State min $\alpha {BCE}_{real}+(1-\alpha)(BCE_{fake}^{local}+BCE_{fake}^{regional})$
    \State Update parameters of $\mathcal{D}$ by minimizing ${\mathcal{G}}_{loss}$:
    \State $\frac{1}{n}\sum{|\mathcal{G} ^{local}(\tilde{X})-X|}^{2}\Rightarrow {MSE}^{local}$
    \State $\frac{1}{n}\sum{|\mathcal{G} ^{regional}(\tilde{X})-X|}^{2}\Rightarrow {MSE}^{regional}$
    \State $BCE(\mathcal{D}(\mathcal{G}^{local}(\tilde{X})),real)\Rightarrow BCE_{real}^{local}$
    \State $BCE(\mathcal{D}(\mathcal{G}^{regional}(\tilde{X})),real)\Rightarrow BCE_{real}^{regional}$
    \State min $(1-\gamma)({MSE}^{local}+{MSE}^{regional})+\gamma(BCE_{real}^{local}+BCE_{real}^{regional})$
    \State Save parameters of $\mathcal{G} ^{local}$, $\mathcal{G} ^{regional}$, $\mathcal{D}$ in current epoch.
    \EndFor
    \For{i $= 0,1,\ldots,len({\mathcal{X}}^{test})$}
    \State Calculate $\mathcal{G} ^{local}$, $\mathcal{G} ^{regional}$ results:
    \State $\mathcal{X}^{test}\Rightarrow {\mathcal{G}}^{local}(\mathcal{X}^{test}),{\mathcal{G}}^{regional}(\mathcal{X}^{test})$
    \State Calculate $\mathcal{D}(X)$ results:
    \State ${\mathcal{D}(\mathcal{G}}^{local}(\mathcal{X}^{test}))$, ${\mathcal{D}(\mathcal{G}}^{regional}(\mathcal{X}^{test}))$ and ${\mathcal{D}(\mathcal{X}^{test})}$
    \State Obtain the combined anomaly score:
    \State $score=\beta({\mathcal{D}(\mathcal{G}}^{local}(\mathcal{X}^{test}))+{\mathcal{D}(\mathcal{G}}^{regional}(\mathcal{X}^{test})))+(1-\beta){\mathcal{D}(\mathcal{X}^{test})}$
    \EndFor
	\end{algorithmic}
\end{algorithm}

%% file: sec/4_experiments.tex
\section{Experiments}
\label{sec: Experiments}
We conduct extensive experiments on four different datasets (three cavitation datasets are provided by SAMSON AG and a public dataset) to validate the effectiveness of the proposed GRLNet. And detailed analysis of the performance and its comparison with state-of-the-art methods \cite{nguyen2021deep,zenati2018efficient,muller2020acoustic,kingma2013auto,baldi2012autoencoders} are also reported. In addition, we provide extensive discussions and ablation experiments to demonstrate the stability and importance of the proposed GRLNet.

\noindent\textbf{Evaluation criteria.} We calculate an anomaly score for each sample $X$, where samples with higher scores are more anomalous. Following previous researches \cite{audibert2020usad,hundman2018detecting,su2019robust}, we select the AUC score of ROC curve area \cite{campos2016evaluation} as an evaluation metric to determine the degree of separability of samples. It measures the one-class classification performance of the model among all possible thresholds and distinguishes between normal and anomaly events. The higher ROC-AUC score, the better the model. A ROC-AUC score which is closer to $1.0$ indicates the model can perfectly differentiate between normal and anomaly events. In addition, we also use F1-score as another evaluation criterion. It represents the average of precision and recall for a specific threshold value. The threshold is determined through Youden index in ROC-AUC (more details see Appendix \ref{sec: AppendixExperiments}).

\subsection{Datasets}
\label{sec: Datasets}
\noindent\textbf{Cavitation Datasets.} This dataset is provided by SAMSON AG and contains three sub-datasets, called cavitation2017, cavitation2018 and cavitation2018 with real noise, respectively. The cavitation acoustic signals are collected from different valves with different upstream pressures and different valve opening rates in a professional environment. The cavitation includes incipient cavitation, constant cavitation and cavitation choked flow. The non-cavitation contains turbulent flow and no flow. Cavitation2017 has a total of 356 acoustic signals and each acoustic signal with time duration of 3 s. Cavitation2018 and Cavitation2018-noise have 806 and 160 acoustic signals and each of them with time duration of 25 s, respectively. The sampling rate of all recording acoustic signals is 1562.5 kHz (more details see Appendix \ref{sec: AppendixExperiments}).

\noindent\textbf{MIMII Dataset.} This dataset \cite{purohit2019mimii} comprises normal and abnormal acoustic signals from four industrial machine types: fan, pump, slider and valve. It records anomalous acoustic signals for different conditions, such as contamination, leakage, clogging, voltage change, rail damage, losses belt, no grease, etc. Each industrial machine type consists of four individual machines, named ID 00, ID 02, ID 04 and ID 06. Considering real industrial noise, each individual machine has three different levels of signal-to-noise ratio (-6 dB, 0 dB and 6 dB) data. The sampling rate for all acoustic signals is 16 kHz and each of them has a time duration of 10 s. The MIMII dataset is split into a training set and a test set. The test set is consisted of all abnormal acoustic signals and the same number of normal acoustic signals. The other normal acoustic signals are regarded as the training set (more details see Appendix \ref{sec: AppendixExperiments}).

\subsection{Anomaly Detection Results}
\label{sec: Experimental Results}
Anomaly detection is one of the significant applications of one-class classification learning algorithm. In unsupervised one-class anomaly detection problem, events belonging to known classes are regarded as normal values for the training of the model. Other events that do not belong to the known classes are treated as anomalous values for model determination. 
\input{fig_input/input_example_signals}

\noindent\textbf{Results on Cavitation datasets.} Figure \ref{fig: examples training and test signals}b shows anomalies examples reconstructed using $\mathcal{G} ^{local}$ and $\mathcal{G} ^{regional}$. It indicates that $\mathcal{G} ^{local}$ and $\mathcal{G} ^{regional}$ can better reconstruct normal signals (non-cavitation) and cannot reconstruct abnormal signals (cavitation). As seen in Table \ref{tab: AUC_cavitation}, GRLNet shows superior results in terms of AUC compared to other state-of-the-art methods. And the proposed GRLNet retains superior and robust performance even with an increased percentage of anomalies at test time. The proposed GRLNet achieves an average of \textbf{0.9855}, \textbf{0.9705} and \textbf{0.9999} AUC in the three cavitation datasets, respectively. 

Detailedly, (1) \noindent\textbf{Cavitation2017 dataset:} The AUC of GRLNet is above \textbf{0.975} for each specific abnormal proportion, which outperforms over all baselines. (2) \noindent\textbf{Cavitation2018 dataset:} The AUC of GRLNet is above \textbf{0.96} as the proportion of abnormalities increases, and it improves on average by \textbf{0.0357} over the standard GAN. (3) \noindent\textbf{Cavitation2018-noise dataset:} Our method obtains perfect performance in all cases with different anomaly proportions. The reasons why our method gets superior results are as follows. First, although this dataset with real background noise compared to other cavitation datasets, the global filter layer can filter most of the noise. Second, this dataset is obtained with only one operating of valve stroke and upstream pressure (see Appendix \ref{sec: AppendixExperiments}).

In addition, we also report ${F}_{1}$ score as an evaluation metric of our method on three real-world cavitation datasets. A comparsion provided in Figure \ref{fig: cavitation f1 results} shows that our approach performs robustly to detect anomalies even when the percentage of anomalies is increased.
\input{tab/Experiments/CavitationDatasets_AUC}
\input{fig_input/input_F1-score_cavitation}

\noindent\textbf{Results on MIMII.} Table \ref{tab: Results-MIMII} shows the experimental results of best AUCs. In general, GRLNet achieves higher AUCs compared to AE, VAE, CAE, CVAE and GAN for each industrial machine type consisting of four individual machines. We observe that most methods achieve high detection performance on fan, pump, slider and valve, but the GRLNet's best-AUC still outperforms them by \textbf{0.0004}-\textbf{0.0188}. 

Detailedly, (1) \noindent\textbf{Fan:} For 0 dB SNR, GRLNet achieves excellent performance for all machines. In addition, GRLNet gets better AUCs than all baselines for ID 02, ID 04 and ID 06 at 6 dB SNR. And AE shows the higher performance at ID 00. For -6 dB SNR, AE and VAE achieve the better performance at ID 02 and ID 04, respectively. The AUCs of GRLNet are improved by 0.0188 and 0.0088 compared to other methods at ID 00 and ID 06, respectively. (2) \noindent\textbf{Pump:} The proposed GRLNet outperforms other methods, except for some certain cases. Specifically, the CVAE obtains the better performance for ID 00 and ID 06 at -6 dB SNR, and ID 00 at 6 dB SNR. Meanwhile, GAN has a higher AUC for ID00 at 0 dB SNR than GRLNet. (3) \noindent\textbf{Slider:} The AE exhibits better performance for ID 04 at both -6 and 6 dB SNRs. CAE shows higher AUCs at ID 04 (0 and 6 dB SNRs). The GAN also has a higher AUC at ID 00 (-6 dB SNR). In other cases, the proposed GRLNet obtains the best performance compared to other baselines, which is improved on average by 0.0332, 0.0484, 0.0415, 0.0203 and 0.0221 compared to AE, VAE, CAE, CAVE and GAN, respectively. (4) \noindent\textbf{Valve:} The performance of the GRLNet and all baselines in valves are significantly lower than other machines due to the instability of valve signals and complex anomalous conditions (more than two kinds of contamination). However, the GRLNet achieves better performance compared to other methods on over half of all machines with different SNRs.

From Table \ref{tab: AVerageResults-MIMII}, it can be seen that the GRLNet gets the highest performance at all SNRs of the pump. In addition, the GRLNet outperforms other methods for 0dB and 6dB SNRs of the fan, -6dB and 6dB SNRs of the slider, and 0dB SNR for the valve. For a few SNR-specific machines, the AVE, CAE and CVAE outperform our method at the best AUC average. 

\noindent\textbf{Takeaways:} The proposed GRLNet achieves the best results for four real-world datasets. This is motivated by two following reasons. Firstly, this ASD task is evaluated based on event anomaly detection of acoustic segments rather than on acoustic frames. Secondly, the global filter layer and the different pattern generators ($\mathcal{G} ^{regional}$ and $\mathcal{G} ^{local}$) make the GRLNet keep both the global spatial acoustic features and the different detailed (local and regional) acoustic features.
\input{tab/Experiments/MIMII}

\input{tab/Experiments/MIMII_Average}

\subsection{Ablation Analysis}
\label{sec: Ablation Analysis}
\noindent\textbf{Detection strategy.} We report different detection strategies in Table \ref{tab: different detection strategies}. It can be seen from Table \ref{tab: different detection strategies} that our detection strategy for GRLNet significantly outperforms the other detection strategies. Where, $\mathcal{D}({\mathcal{G}}^{local}(X)) + \mathcal{D}({\mathcal{G}}^{regional}(X))$ and $\mathcal{D}(X)$ and  are special cases of our detection strategy $\beta \mathcal{D}({\mathcal{G}}^{local}(X)) + \mathcal{D}({\mathcal{G}}^{regional}(X)) + (1-\beta)\mathcal{D}(X)$ when $\beta=1.0$ and $\beta=0$, respectively.
\input{tab/Experiments/DetectionStrategy}
\input{fig_input/input_ablationanalysis}

\noindent\textbf{Analysis of window size.} We compare the anomaly detection results of our method with different window sizes in cavitation2017 (anomaly percentage of 50$\%$). It is clear from Figure \ref{fig: ablation analysis}a that the window size is sensitive to the detection results of GRLNet.

\noindent\textbf{Parameter sensitivity.} We show the impact of different values of the comprehensive output factor $\beta$ on the detection results of the proposed GRLNet in cavitation2017 (anomaly percentage of 50$\%$), see \ref{fig: ablation analysis}b. When $\beta\in(0,1)$, the GRLNet is not sensitive to $\beta$ and the detection strategy is $\mathcal{D}({\mathcal{G}}^{local}(X)) + \mathcal{D}({\mathcal{G}}^{regional}(X)) + \mathcal{D}(X)$ in this case. In general, we set $\beta=0.5$. Where, the detection strategy is $\mathcal{D}(X)$ when $\beta=0$; when $\beta=1.0$, the detection strategy is $\mathcal{D}({\mathcal{G}}^{local}(X)) + \mathcal{D}({\mathcal{G}}^{regional}(X))$.

\noindent\textbf{Analysis of downsampling.} In practical applications, the resolution of sensors represent the quality of the data obtained. The ability to detect different levels of the data obtained from low-level sensors becomes very significant and challenging. Our method is evaluated in cavitation2017 (anomaly percentage of 50$\%$) under the original frequency of samples ($Fs$ = 1562500 Hz), one-half, one-quarter, one-sixth, one-eighth of the original frequency of samples (781250 Hz, 390625 Hz, 260416 Hz and 195312 Hz) and that of the mobile phone can mostly accommodate (48000 Hz $\approx Fs/32$), respectively. As shown in Figure \ref{fig: ablation analysis}c, although the proposed GRLNet begins to suffer from a gradual reduction in AUC and ${F}_{1}$ score as the frequency of samples decreases, the AUC and ${F}_{1}$ score always remain above \textbf{0.96} when 1562500 Hz $\geq  Fs\geq$ 195312 Hz. And even for mobile phone setup with the frequency of samples equals to 48000 Hz, the proposed GRLNet achieves an AUC and ${F}_{1}$ score of \textbf{0.9611} and \textbf{0.9655}, respectively.

\noindent\textbf{Individual abnormal state detection capability.} We compare the detection capability of GRLNet in individual cavitation states in Cavitation2017 (anomaly percentage of 50$\%$), i.e., the anomalies all consist of a individual cavitation state. It can be seen from Figure \ref{fig: ablation analysis}d that both continuous and choked cavitation can be detected well. However, the incipient cavitation is not well detected since it is the critical state between cavitation and non-cavitation.

%% file: fig_input/input_example_signals.tex
\begin{figure}[htbp]
\centering
\subfigure[Training signals]{
\begin{minipage}[t]{\linewidth}
\centering
\includegraphics[width=\textwidth,height=40mm]{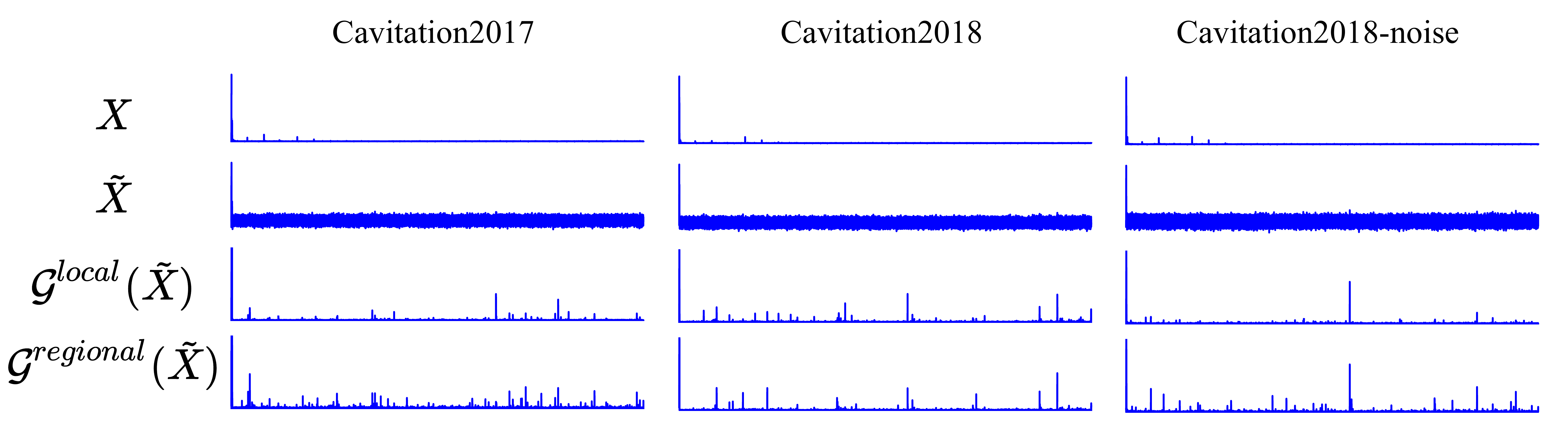}
\end{minipage}%
}%

\subfigure[Test signals]{
\begin{minipage}[t]{\linewidth}
\centering
\includegraphics[width=\textwidth,height=55mm]{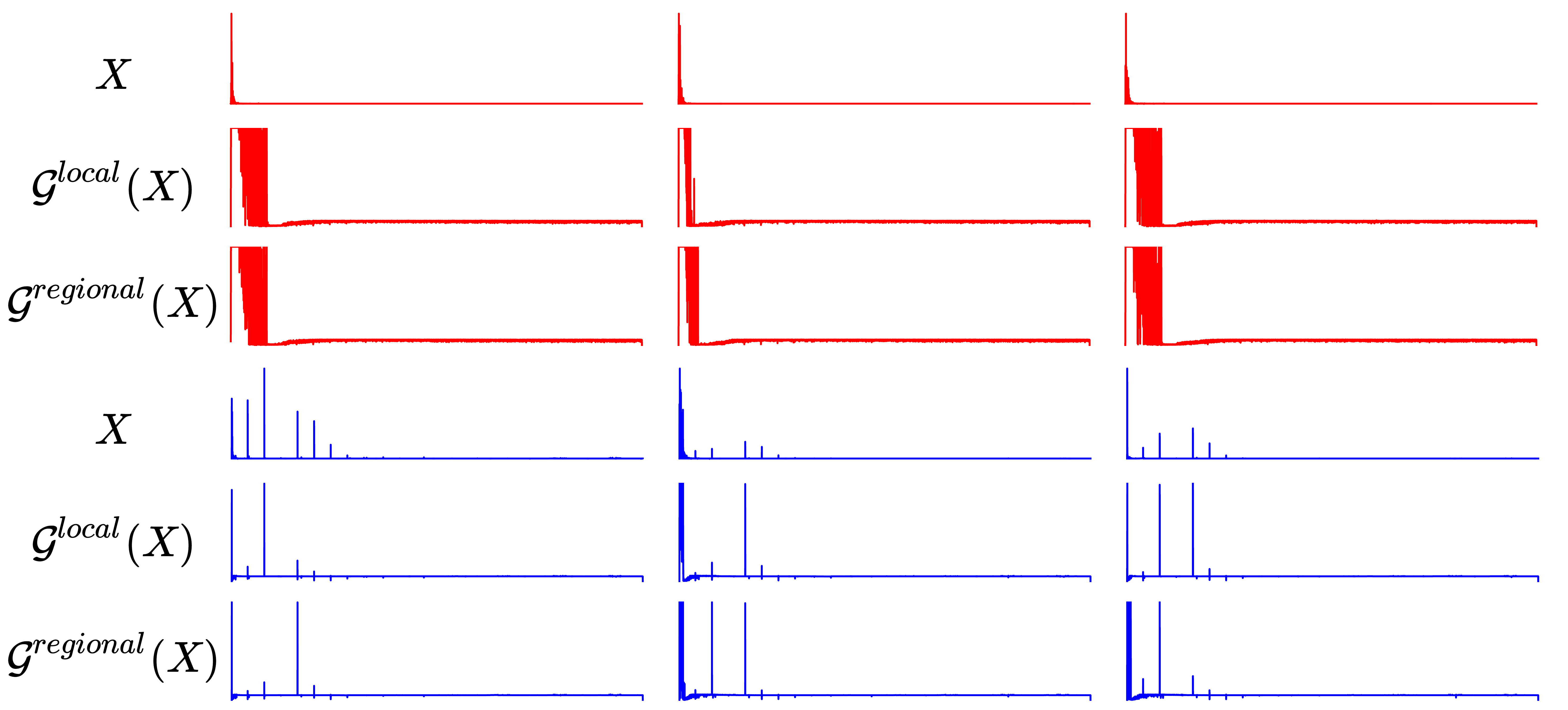}
\end{minipage}%
}%
\centering
\caption{Example acoustic signals from different stages of our framework (Frequency (kHz) as the $x$-axis and normalized amplitude as the $y$-axis). From top to bottom: The output signals from global filter layer $\bm{X}$, ${\tilde X}$ shows the signal with noise, local pattern reconstructed ${\mathcal{G}}^{local}(\cdot)$ and regional pattern reconstructed ${\mathcal{G}}^{regional}(\cdot )$. Blue and red colors represent normal and abnormal, respectively.}
\label{fig: examples training and test signals}
\end{figure}

%% file: tab/Experiments/CavitationDatasets_AUC.tex
\begin{table}[htbp]
\caption{AUC results on three real-world cavitation datasets. The window size is 466944 and step size is 466944. }
\label{tab: AUC_cavitation}
\footnotesize
\setlength{\tabcolsep}{0.4mm}{
\begin{tabular}{lccccc|ccccc|ccccc}
\toprule
\multicolumn{1}{c}{Dataset}                     & \multicolumn{5}{c}{Cavitation2017} & \multicolumn{5}{c}{Cavitation2018} & \multicolumn{5}{c}{Cavitation2018-noise} \\
\midrule
\multicolumn{1}{c}{Abnormal percentage}& 10     & 20     & 30     & 40     & 50     & 10     & 20     & 30      & 40     & 50   & 10     & 20     & 30     & 40    & 50    \\ 
\midrule
AE \cite{baldi2012autoencoders}        &0.8981 &0.8871 &0.8862 &0.8821 &0.8816  &0.8881 &0.8876 &0.8624 &0.8549 &0.8483  &0.9378 &0.9346 &0.9282 &0.9218 &0.9178       \\
VAE \cite{kingma2013auto}              &0.9398 &0.9387 &0.9340 &0.9290 &0.9152  &0.9383 &0.9319 &0.9299 &0.9159 &0.9158  &0.9498 &0.9474 &0.9442 &0.9430 &0.9424    \\
GAN \cite{zenati2018efficient}         &0.9449 &0.9424 &0.9382 &0.9365 &0.9262  &0.9462	&0.9419	&0.9283	&0.9299	&0.9276  &0.9578 &0.9524 &0.9510 &0.9499 &0.9499       \\
GRLNet& \textbf{1.0000} &\textbf{0.9901} &\textbf{0.9829} &\textbf{0.9790} &\textbf{0.9756}  &\textbf{0.9918}	&\textbf{0.9716} &\textbf{0.9634}&\textbf{0.9633}&\textbf{0.9622} &\textbf{1.0000} 	&\textbf{0.9999} &\textbf{0.9999} &\textbf{0.9999}&\textbf{0.9999}             \\ 
\bottomrule
\end{tabular}
}
\end{table}

%% file: fig_input/input_F1-score_cavitation.tex
\begin{figure*}[htbp]
\centering
\subfigure[Cavitation2017]{
\begin{minipage}[t]{0.33\linewidth}
\centering
\includegraphics[width=\textwidth,height=35mm]{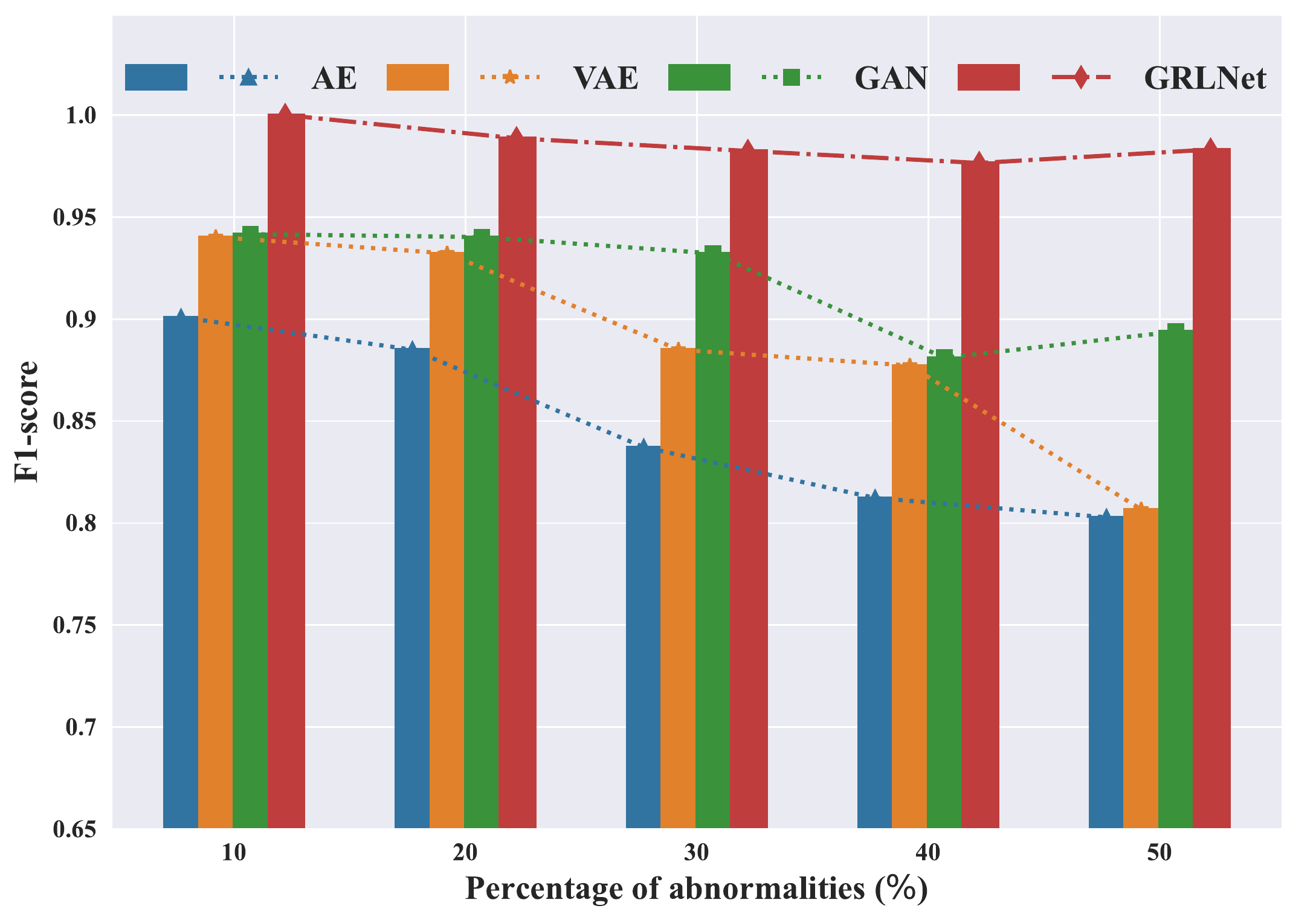}
\end{minipage}%
}%
\subfigure[Cavitation2018]{
\begin{minipage}[t]{0.33\linewidth}
\centering
\includegraphics[width=\textwidth,height=35mm]{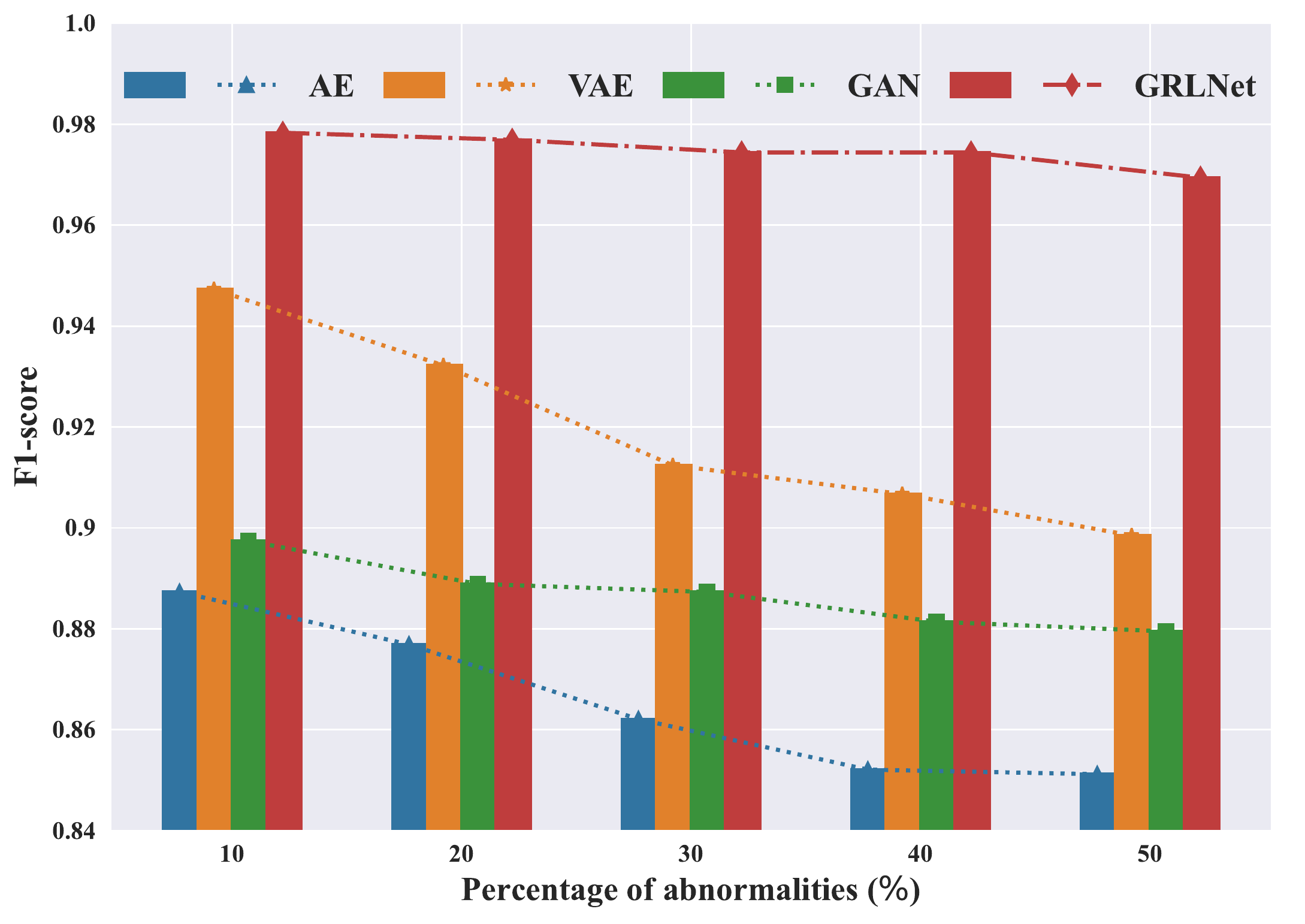}
\end{minipage}%
}%
\subfigure[Cavitation2018-noise]{
\begin{minipage}[t]{0.33\linewidth}
\centering
\includegraphics[width=\textwidth,height=35mm]{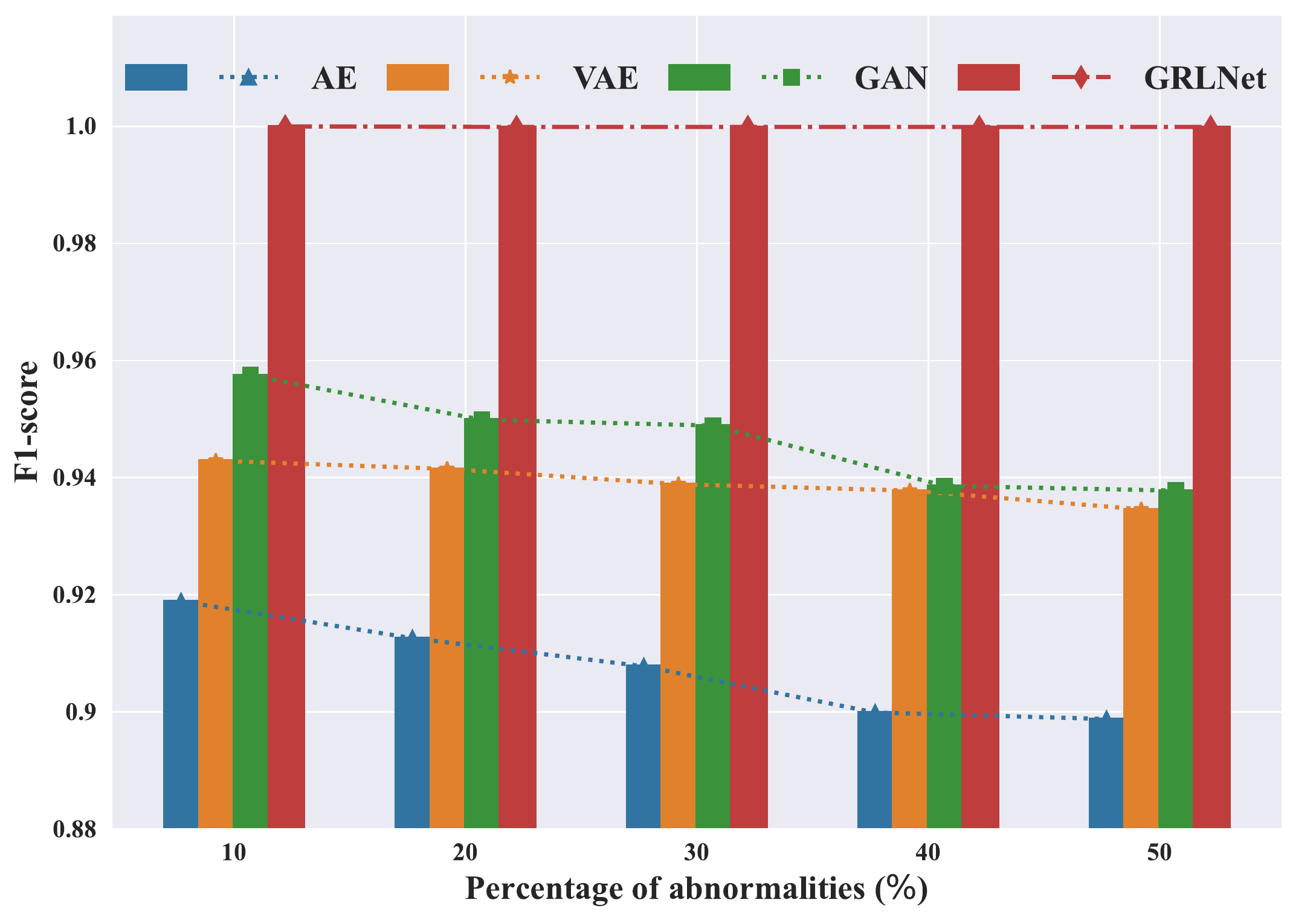}
\end{minipage}%
}%
\centering
\caption{${F}_{1}$ score results on three real-world cavitation datasets. The window size is 466944 and step size is 466944.} 
\label{fig: cavitation f1 results}
\end{figure*}

%% file: tab/Experiments/MIMII.tex

\begin{table}[htbp]
\centering
\caption{Comparison with anomaly detection architectures on MIMII. We report the best AUC for each industrial machine type (Fan, Pump, Slider and Valve) consisting of four individual machines (ID 00, ID 02, ID 04 and ID 06) with three different levels of signal-to-noise ratio (-6 dB, 0 dB and 6 dB).}
\label{tab: Results-MIMII}
\footnotesize
\setlength{\tabcolsep}{1mm}{
\begin{tabular}{clcccc|cccc|cccc}
\toprule 
\multicolumn{2}{c}{Input SNR}   & \multicolumn{4}{c}{-6 dB}         & \multicolumn{4}{c}{0 dB}          & \multicolumn{4}{c}{6 dB}           \\  
\midrule
Machine                  & Model                           & 00     & 02     & 04     & 06     & 00     & 02     & 04     & 06     & 00     & 02     & 04     & 06      \\  
\midrule
\multirow{5}{*}{Fan}& AE \cite{baldi2012autoencoders} &0.5629 &\textbf{0.7182} &0.5957 &0.8723  & 0.6414 & 0.9266 & 0.8327 & 0.9932     &\textbf{0.8559} & 0.9924 & 0.9635 & \textbf{1.0000} \\
                    & VAE \cite{kingma2013auto}       &0.5676 &0.7054 &\textbf{0.7054} &0.8802  & 0.6771 & 0.9297 & 0.7942 & 0.9943       & 0.8285 & 0.9949 & 0.9671 & \textbf{1.0000 }\\
                    & CAE \cite{muller2020acoustic}   &0.5462 &0.6341 &0.5839 &0.7452           & 0.5062 & 0.7762 & 0.7373 & 0.9230       & 0.6773 & 0.9322 & 0.8190 & 0.9912 \\
                    & CVAE \cite{nguyen2021deep}      &0.5349 &0.6754 &0.5405 &0.7899           & 0.6115 & 0.8143 & 0.7452 & 0.9799       & 0.7313 & 0.9844 & 0.8949 & 0.9994 \\
                    & GAN \cite{zenati2018efficient}  &0.5417 &0.6639 &0.6125 &0.8826           &0.6538 &0.9188 &0.8315 & 0.9829 & 0.8147 & 0.9869 & 0.9615 & 0.9943 \\
& GRLNet &\textbf{0.5864} &0.6874 &0.6321 &\textbf{0.8914}  &\textbf{0.6857} &\textbf{0.9468} &\textbf{0.8375}&\textbf{0.9949} &0.8359   &\textbf{0.9996}   &\textbf{0.9781} &\textbf{1.0000}\\
\midrule
\multirow{5}{*}{Pump}   & AE \cite{baldi2012autoencoders} & 0.7425 & 0.6220 & 0.9215 & 0.5539      & 0.7646 & 0.6176 & 0.9696 & 0.9127       & 0.9516 & 0.5816 & 0.9991 & 0.9423 \\
        & VAE \cite{kingma2013auto}       & 0.7289 & 0.6080 & 0.9168 & 0.5849      & 0.7740 & 0.6289 & 0.9699 & 0.8945       & 0.9227 & 0.5991 &\textbf{1.0000}& 0.9856 \\
        & CAE \cite{muller2020acoustic}   & 0.7574 & 0.7329 & 0.8457 & 0.6092      & 0.7335 & 0.6790 & 0.9076 & 0.7636       & 0.8522 & 0.6007 & 0.9937 & 0.9846 \\
        & CVAE \cite{nguyen2021deep}& \textbf{0.7959} & 0.6276 & 0.9369 & \textbf{0.6180} & 0.8252 & 0.6366 & 0.9743 & 0.7892       &\textbf{0.9983} & 0.5578 &\textbf{1.0000}& 0.9926 \\
        & GAN \cite{zenati2018efficient}  & 0.7759 & 0.7048 & 0.9399 & 0.5979 & \textbf{0.8386} & 0.6792 & 0.9699 & 0.8946 & 0.9637   & 0.6097   & 0.9847   & 0.9786 \\
& GRLNet & 0.7845 &\textbf{0.7480}& \textbf{0.9479}& 0.6179   & 0.8286   & \textbf{0.6796} & \textbf{0.9896} & \textbf{0.9147}& 0.9887   & \textbf{0.6216}&\textbf{1.0000}&\textbf{0.9946} \\
\midrule
\multirow{5}{*}{Slider} & AE \cite{baldi2012autoencoders} & 0.9613 & 0.7514 &\textbf{0.6838} & 0.5403 & 0.9905 & 0.8186 & 0.8105 & 0.5200 & 0.9938 & 0.8908 &\textbf{0.9203} & 0.8066 \\
        & VAE \cite{kingma2013auto}       & 0.9573 & 0.7062 & 0.6627 & 0.4955      & 0.9893 & 0.8106 & 0.7851 & 0.5395       & 0.9945 & 0.9084 & 0.9163 & 0.7702 \\
        & CAE \cite{muller2020acoustic}   & 0.9704 & 0.6785 & 0.5610 & 0.6362      & 0.9958 & 0.8236 & 0.8110 &\textbf{0.6805} &\textbf{0.9996} & 0.7936 & 0.7410 &\textbf{0.8472}\\
        & CVAE \cite{nguyen2021deep}      & 0.9740 & 0.7770 & 0.5948 & 0.5215      & 0.9987 & 0.8294 &0.8438 & 0.5016       &\textbf{0.9996} & 0.9166 & 0.8593 & 0.8021 \\
        & GAN \cite{zenati2018efficient}  & \textbf{0.9773} & 0.7648 & 0.5847 & 0.5615 & 0.9953 & 0.8241 & 0.8128 & 0.5135  & 0.9943 & 0.9048 & 0.8493 & 0.8182\\
& GRLNet & 0.9742 &\textbf{0.7884} & 0.5928 & \textbf{0.6435} & \textbf{0.9991} & \textbf{0.8321} & \textbf{0.8451} & 0.5575 &\textbf{0.9996} &\textbf{0.9208} & 0.8873 & 0.8362 \\ 
\midrule
\multirow{5}{*}{Valve}  & AE \cite{baldi2012autoencoders} & 0.4529 & 0.5539 & 0.5134 & 0.4992 & 0.4943 & 0.5932 & 0.5569 & 0.5517       & 0.5248 & 0.6940 & 0.5701 & 0.5964 \\
        & VAE \cite{kingma2013auto}       & 0.4511 & 0.5236 & 0.5154 & 0.5013      & 0.4880 & 0.5903 & 0.5738 & 0.5347       & 0.5087 & 0.6440 & 0.5424 & 0.5964 \\
        & CAE \cite{muller2020acoustic}   & 0.5015 & 0.5743 & 0.5235 & 0.5041      & 0.3996 & 0.5874 &\textbf{0.6081} & 0.5310       & 0.3477 & 0.6006 & 0.5410 & 0.5487 \\
        & CVAE \cite{nguyen2021deep}      & 0.5132 & 0.5865 &\textbf{0.5253} &\textbf{0.5465} & 0.5153 &\textbf{0.6367} &0.5570 &0.5699 &\textbf{0.6584} &0.7394 &0.5875 &\textbf{0.6342} \\
        & GAN \cite{zenati2018efficient}  & 0.5065 & 0.5686 & 0.5185 & 0.5025 & 0.4813 & 0.5736 & 0.5649 & 0.5517  & 0.5297 & 0.7128 & 0.5501   & 0.5964 \\
& GRLNet &\textbf{0.5219}&\textbf{0.5875} & 0.5154 & 0.5115 &\textbf{0.5156} & 0.6062 & 0.5859 &\textbf{0.5727} & 0.6137 & \textbf{0.7426} &\textbf{0.5924} & 0.6084 \\ 
\bottomrule
\end{tabular}
}
\end{table}

%% file: tab/Experiments/MIMII_Average.tex
\begin{table}[htbp]
\centering
\caption{Comparison of best AUC averages on MIMII.}
\label{tab: AVerageResults-MIMII}
\footnotesize
\setlength{\tabcolsep}{1mm}{
\begin{tabular}{cccccccc}
\toprule
\multirow{2}{*}{Machine} & \multirow{2}{*}{\begin{tabular}[c]{@{}c@{}}Input \\ SNR\end{tabular}} & \multicolumn{6}{c}{Model}\\ 
\cmidrule{3-8} 
                         &                & AE              & VAE                    & CAE              & CVAE             & GAN           & GRLNet            \\ 
\midrule
\multirow{3}{*}{Fan}     & -6 dB          & 0.6873          & \textbf{0.7147}        & 0.6274           & 0.6352           & 0.6752        & 0.6993           \\
                         & 0 dB           & 0.8485          & 0.8488                 & 0.7357           & 0.7877           & 0.8468        & \textbf{0.8662}  \\
                         & 6 dB           & 0.9530          & 0.9476                 & 0.8549           & 0.9025           & 0.9394        &\textbf{0.9534} \\ 
\midrule
\multirow{3}{*}{Pump}    & -6 dB          & 0.7100          & 0.7097                 & 0.7363           & 0.7446           & 0.7546        & \textbf{0.7746}   \\
                         & 0 dB           & 0.8161          & 0.8168                 & 0.7709           & 0.8063           & 0.8456        & \textbf{0.8531}  \\
                         & 6 dB           & 0.8686          & 0.8769                 & 0.8578           & 0.8872           & 0.8842        & \textbf{0.9012}  \\ 
\midrule
\multirow{3}{*}{Slider}  & -6 dB          & 0.7342          & 0.7054                 & 0.7115           & 0.7168           & 0.7221        & \textbf{0.7497}  \\
                         & 0 dB           & 0.7849          & 0.7811                 & \textbf{0.8277}  & 0.7934           & 0.7864        & 0.8085            \\
                         & 6 dB           & 0.9027          & 0.8974                 & 0.8454           & 0.8944           & 0.8917        & \textbf{0.9110}   \\ 
\midrule
\multirow{3}{*}{Valve}   & -6 dB          & 0.5026          & 0.4979                 & 0.5259           & \textbf{0.5429}  & 0.5240        & 0.5341            \\
                         & 0 dB           & 0.5486          & 0.5467                 & 0.5315           & 0.5697           & 0.5429        & \textbf{0.5701}    \\
                         & 6 dB           & 0.5947          & 0.5729                 & 0.5095           & \textbf{0.6549}  & 0.5973        & 0.6393            \\
\bottomrule
\end{tabular}
}
\end{table}

%% file: tab/Experiments/DetectionStrategy.tex
\begin{table}[htbp]
\centering
\caption{Comparison of different detection strategies.}
\label{tab: different detection strategies}
\footnotesize
\setlength{\tabcolsep}{0.1mm}{
\begin{tabular}{ccccc}
\toprule 
\multirow{2}{*}{Detection Strategies}       & \multicolumn{2}{c}{Type}  & \multicolumn{2}{c}{Results} \\ 
\cmidrule{2-5}                              & $G(\cdot)$     & $D(\cdot)$           & AUC         & F1-score      \\ 
\midrule
$\mathcal{D}(X)$                            & \ding{55}     & \ding{51}             & 0.6333      & 0.7317        \\
$\mathcal{G}^{local}(X)$                    & \ding{51}     & \ding{55}             & 0.5167      & 0.6667        \\
$\mathcal{G}^{regional}(X)$                 & \ding{51}     & \ding{55}             & 0.5258      & 0.6667        \\
$\mathcal{D}({\mathcal{G}}^{local}(X))$     & \ding{51}     & \ding{51}             & 0.7859      & 0.8067        \\
$\mathcal{D}({\mathcal{G}}^{regional}(X))$  & \ding{51}     & \ding{51}             & 0.8356      & 0.8487        \\
$\mathcal{D}({\mathcal{G}}^{local}(X)) + \mathcal{D}({\mathcal{G}}^{regional}(X))$           & \ding{51}     & \ding{51}             & 0.9570      & 0.9655        \\
$\beta(\mathcal{D}({\mathcal{G}}^{local}(X)) + \mathcal{D}({\mathcal{G}}^{regional}(X))) + (1-\beta)\mathcal{D}(X)$ & \ding{51}& \ding{51}&\textbf{0.9756}&\textbf{0.9831}\\ 
\bottomrule
\end{tabular}
}
\end{table}

%% file: fig_input/input_ablationanalysis.tex
\begin{figure}[htbp]
\centering
\subfigure[Analysis of window size]{
\begin{minipage}[t]{0.24\linewidth}
\centering
\includegraphics[width=\textwidth,height=26mm]{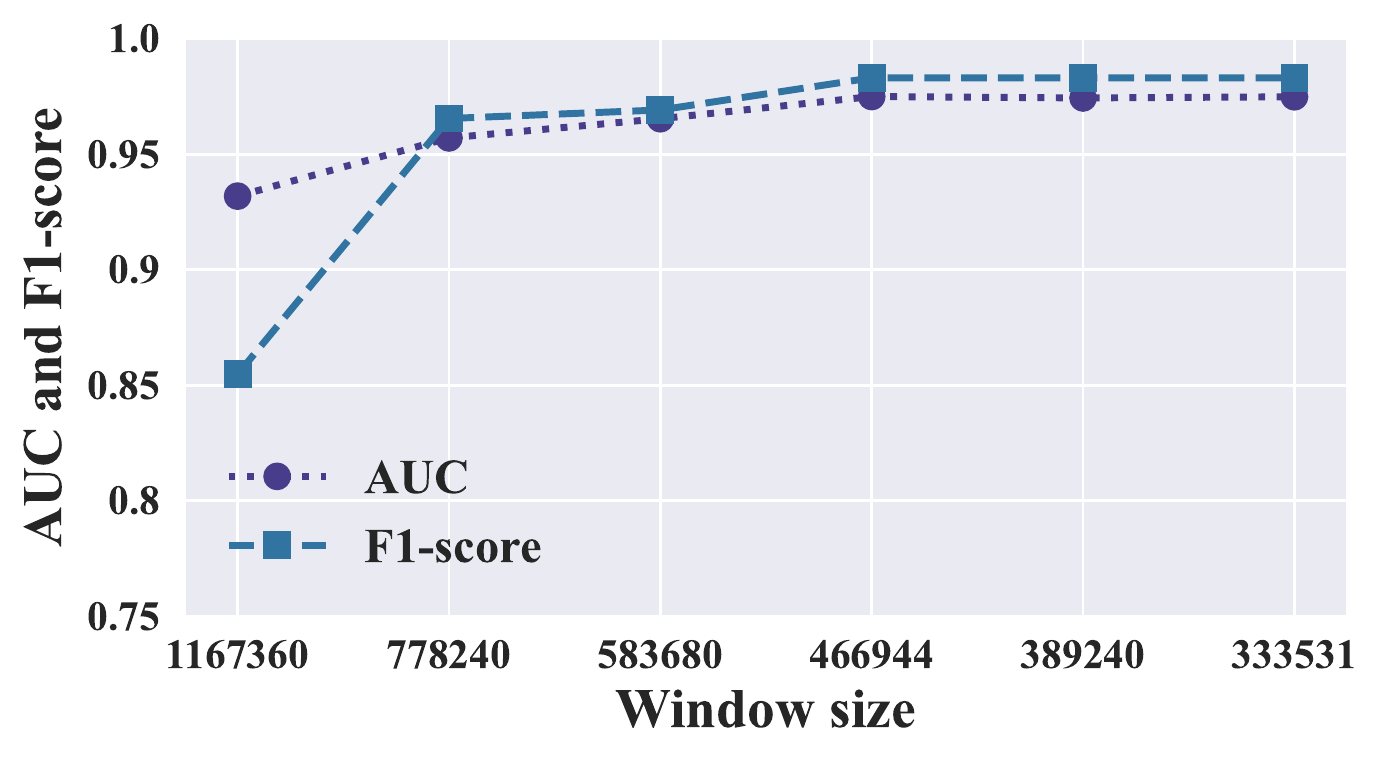}
\end{minipage}%
}%
\subfigure[Parameter sensitivity]{
\begin{minipage}[t]{0.24\linewidth}
\centering
\includegraphics[width=\textwidth,height=26mm]{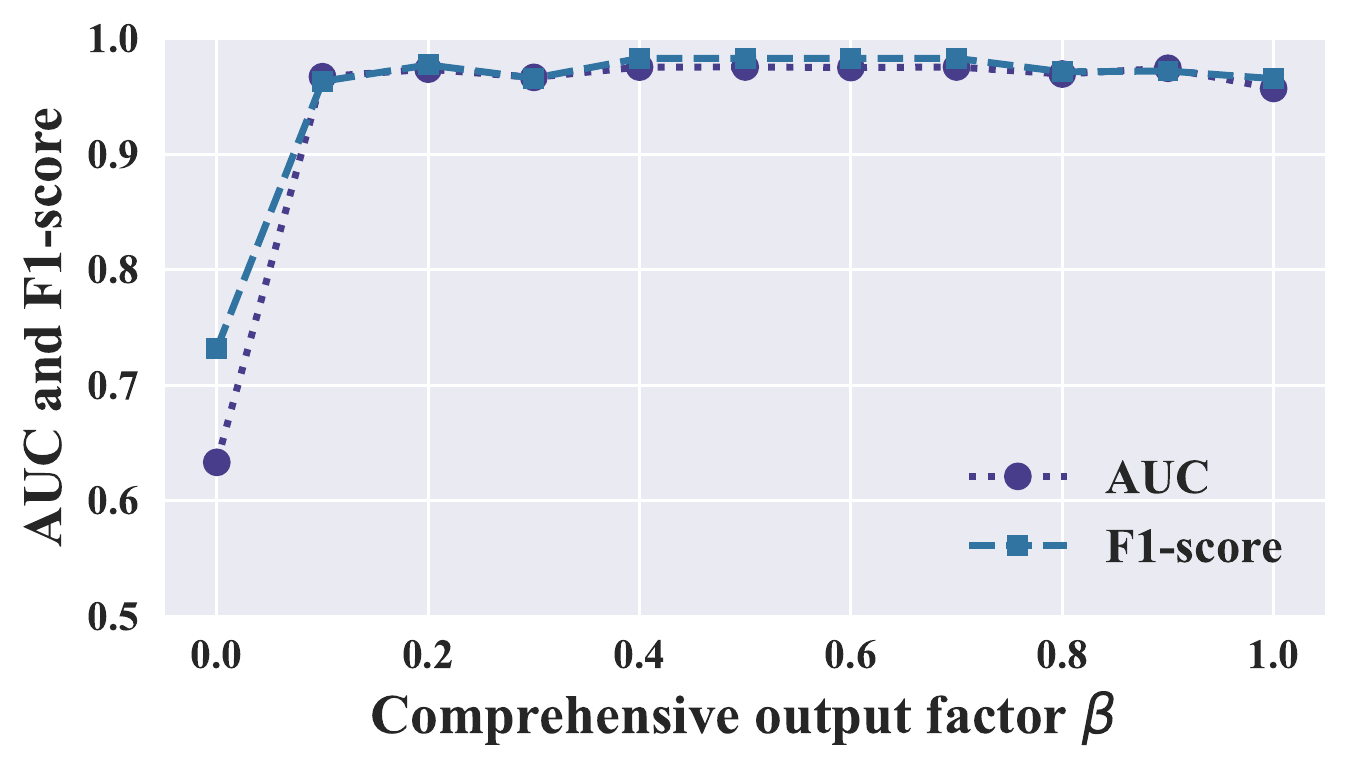}
\end{minipage}%
}%
\subfigure[Analysis of downsampling]{
\begin{minipage}[t]{0.24\linewidth}
\centering
\includegraphics[width=\textwidth,height=26mm]{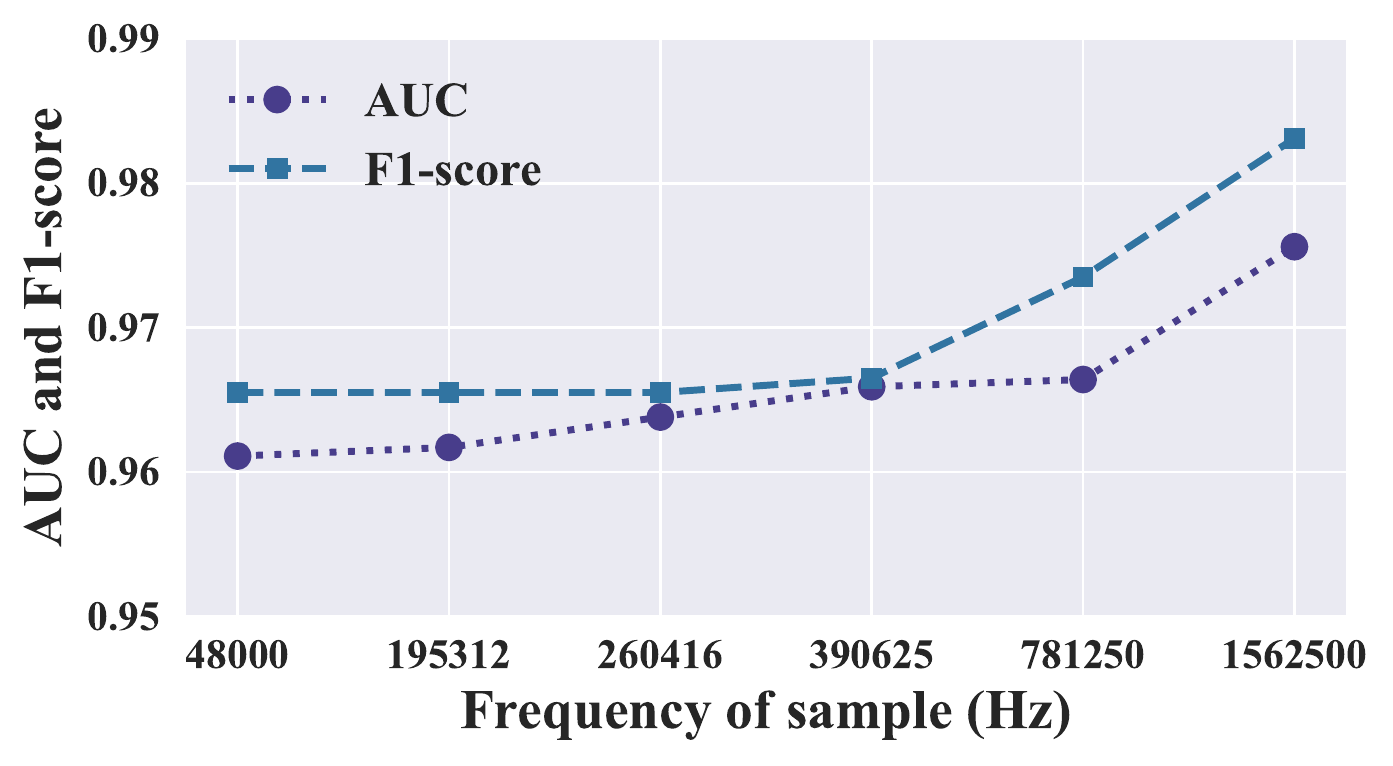}
\end{minipage}%
}%
\subfigure[Individual abnormal state]{
\begin{minipage}[t]{0.24\linewidth}
\centering
\includegraphics[width=\textwidth,height=26mm]{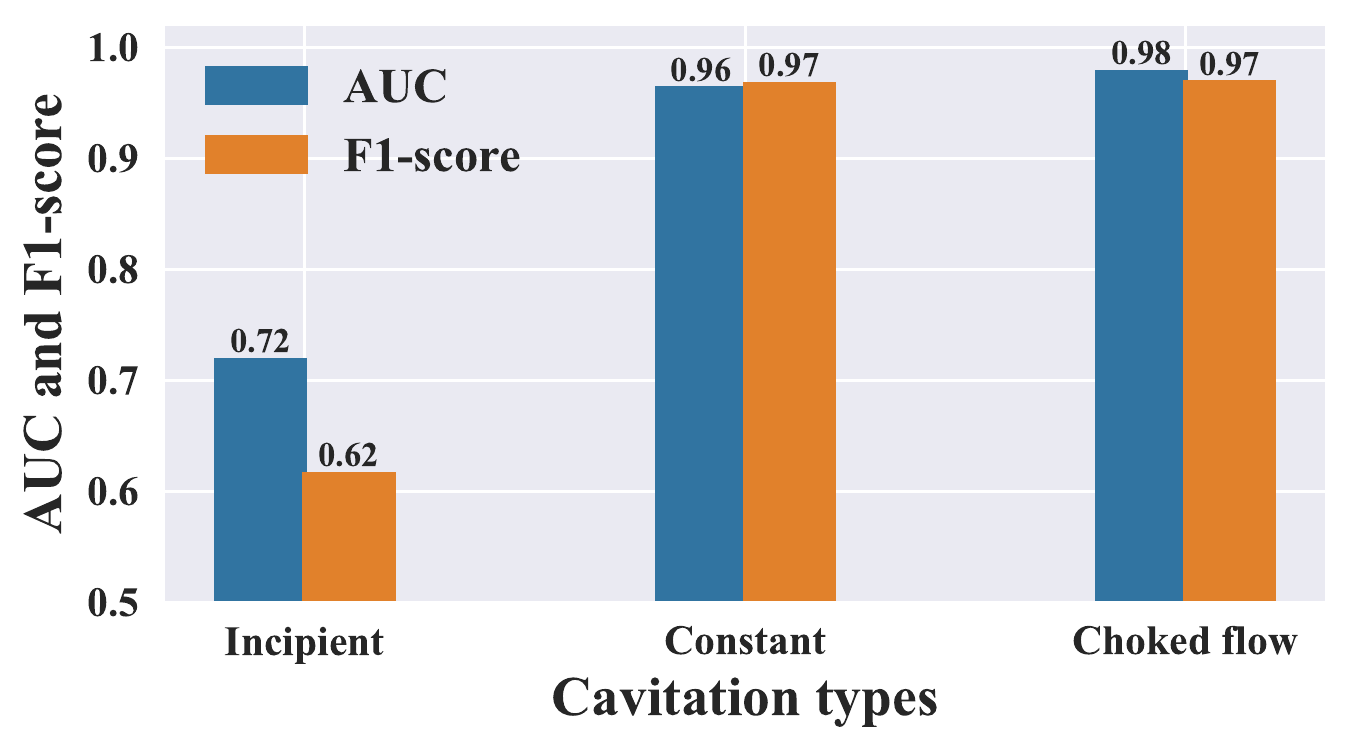}
\end{minipage}%
}%
\centering
\caption{AUC and ${F}_{1}$ score results of various ablation experiments. (a)-(c) are all experimented in Cavitation2017 with anomaly percentage of 50$\%$.} 
\label{fig: ablation analysis}
\end{figure}

%% file: sec/5_relatedwork.tex
\section{Related Work}
\label{sec: related work}
\noindent\textbf{Anomaly Detection.} Anomaly detection is often regarded as a novelty detection problem in which is trained based on a know normal class and ultimately unknown outliers as abnormal. Data-based regeneration works by using generative networks to learn features in an unsupervised manner. Zhou et al. in \cite{zhou2017anomaly} proposed to use deep autoencoder to learn the reconstruction of normal images. Lonescu et al. in \cite{ionescu2019object} to use convolutional auto-encoders based on object detection to learn motion and appearance representations. Su et al. in \cite{su2019robust} proposed a multivariate time series anomaly detection based on stochastic recurrent neural networks. Xu et al. in \cite{xu2017detecting} used a one-class SVM using features from stacked auto-encoders. Ravanbakhsh et al. \cite{ravanbakhsh2017abnormal} used generators as a reconstructor to detect anomalous events, assuming that the generator is unable to reconstruct inputs that do not match the normal training data. Pathak et al. \cite{pathak2016context} proposed adversarial training to improve the quality of regeneration. However, they also discarded the discriminator at the end of training.

\noindent\textbf{Fourier transform in network.} Fourier transform has been an important tool for digital signal and image processing for decades \cite{madisetti1997digital,pitas2000digital}. With the breakthroughs of CNNs in signal and vision, there are various works that start to incorporate Fourier transform into deep learning methods. Some of these studies employ discrete Fourier transform to convert images to the frequency domain and leverage frequency information to improve performance on certain tasks \cite{rao2021global,yang2020fda}. There are also some studies that use the convolution theorem of the Fourier transform to accelerate CNNs \cite{li2020falcon}.

%% file: sec/6_conclusions.tex
\section{Conclusions}
ASD is an essential task for complex industrial system monitoring. In this paper, we propose GRLNet, a novel unsupervised anomaly detection method, which tackles global long-term interactions using a global filter layer and captures the normal patterns of acoustic signals employing multi-pattern generators. Furthermore, we extend the fundamental role of a discriminator from identifying real and fake data to distinguishing between regional and local reconstructions. Moreover, we use both generators and discriminators for acoustic anomaly detection, which breaks the tradition of using only generator and discriminator for acoustic anomaly detection. GRLNet outperforms the SOTA methods on four real-world dataset. The ablation experiments clearly show that GRLNet successfully meets the anomaly detection requirements of complex industrial systems in the real world, which provides insights in applying the method on other industrial applications.

%% file: sec/Acknowledgements.tex
\section*{Acknowledgements}
\label{sec:acknowledgements}
This research is supported by Xidian-FIAS International Joint Research Center (Y. S.), by the AI grant at FIAS through SAMSON AG (J. F., K. Z.),  by the BMBF funding through the ErUM-Data   project (K. Z.), by SAMSON AG (D. V., T. S., A. W.), by the Walter GreinerGesellschaft zur F\"orderung der physikalischen Grundla-genforschung e.V. through the Judah M. Eisenberg Lau-reatus Chair at Goethe Universit\"at Frankfurt am Main (H. S.), by the NVIDIA GPU grant through NVIDIA Corporation (K. Z.).

%% file: sec/supplementary.tex
\setcounter{table}{0}
\setcounter{figure}{0}
\setcounter{equation}{0}
\renewcommand{\thetable}{A\arabic{table}}
\renewcommand{\thefigure}{A\arabic{figure}}
\renewcommand{\theequation}{A.\arabic{equation}}

\section{Discrete Fourier transform}
\label{sec: Discrete Fourier transform}
\subsection{From continuous Fourier transform to discrete Fourier transform}
\noindent Discrete Fourier transforms (DFT) can be derived in many ways. Since the Fourier transform (FT) is originally designed for continuous signals. Therefore, we will derive the DFT from the one-dimensional continuous Fourier transform (CFT). Specifically, given a continuous time signal $f(t)$, 1D CFT is given by:
\begin{small}
\begin{equation}
\label{eq: CFT}
F(u)=\int^{+\infty }_{-\infty }f(t){e}^{-2j\pi ut}dt
\end{equation}
\end{small}
The continuous function $f(t)$ is sampled once at a time interval $\Delta T$ from a certain time (noted as moment $0$) and a total of N times. Then, we get a discretised sequence:
\begin{small}
\begin{equation}
\label{eq: discretised sequence1}
f(\Delta T),f(2\Delta T),\ldots ,f(N\Delta T)
\end{equation}
\end{small}
and the discrete sequence is denoted as:
\begin{small}
\begin{equation}
\label{eq: discretised sequence2}
\hat{f}(0),\hat{f}(1),\ldots ,\hat{f}(N-1)
\end{equation}
\end{small}
where for the other unsampled points, i.e. $\forall t\notin \left \{\Delta T,2\Delta T,\ldots ,N\Delta T\right \}$, the value of the function for these points are $0$, i.e. $f(t)=0$. Applying equations \ref{eq: CFT} and \ref{eq: discretised sequence1}, we have
\begin{small}
\begin{equation}
\begin{aligned}
F(u)&=\int^{+\infty}_{-\infty}f(t){e}^{-2j\pi ut}dt\\
&=\int^{0}_{-\infty}+\int^{N\Delta T}_{0}+\int^{+\infty}_{N\Delta T}f(t){e}^{-2j\pi ut}dt\\
&=\int^{0}_{-\infty}+\int^{+\infty}_{N\Delta T}f(t){e}^{-2j\pi ut}dt+\int^{N\Delta T}_{0}f(t){e}^{-2j\pi ut}dt\\
&=0+\int^{N\Delta T}_{0}f(t){e}^{-2j\pi ut}dt\\
&=\int^{N\Delta T}_{0}f(t){e}^{-2j\pi ut}dt\\
&=F'(u)
\end{aligned}
\end{equation}
\end{small}
Next, we divide the interval $\left [0,N\Delta T\right]$ equally into N parts, the $i$-th interval ${k}_{i}=\left [\left(i-1\right)\Delta T,i\Delta T\right],i=1,2,\ldots ,N$ and the length of each interval is $N$. Let the value of the function taking the right endpoint on each interval ${k}_{i}$ be $f(i\Delta T){e}^{-2j\pi u\cdot (i\Delta T)}$. Furthermore, we can convert $F'(u)$ to the Riemann sum form of $F''(u)$ as follows
\begin{small}
\begin{equation}
\begin{aligned}
F''(u) &= \sum\limits_{i = 1}^N {\left[f\left(i\Delta T\right){e^{ - 2j\pi u \cdot \left(i\Delta T\right)}}\right]}\cdot \Delta T \\
&=\sum_{n=0}^{N-1}\left [f\left( \left(n+1\right)\Delta T\right){e}^{-2j\pi u\cdot\left(n+1\right)\Delta T}\right]\cdot\Delta T\\
&=\sum_{n=0}^{N-1}\left [ \hat{f}\left(n\right){e}^{-2j\pi u\cdot \left(n+1\right)\Delta T}\right]\cdot\Delta T
\end{aligned}
\end{equation}
\end{small}
If we let the length between the start and end times of sampling be 1 time unit, i.e. $N\Delta T=1$, then we have:
\begin{small}
\begin{equation}
F''(u) = \frac{1}{N}\sum\limits_{n = 0}^{N - 1} {\hat f(n){e^{ - 2j\pi u\frac{{n + 1}}{N}}}} ,u = 0,1, \ldots ,N - 1
\end{equation}
\end{small}
which is exactly the formulation of the DFT.

\subsection{The convolvational theorem}
\noindent The convolution theorem is the main property of the Fourier transform. Specifically, the FT for the circular convolution of two discrete sequences is equivalent to the dot product of these two sequences in the frequency domain. Given two discrete sequences $x\left [n\right]$ (signals) and $h\left [n\right]$ (filters), both of length $N$, the circular convolution can be defined as:
\begin{small}
\begin{equation}
w\left [n\right]={\left(h\ast x\right)}_{n}=\sum_{m=0}^{N-1}h\left [m\right]x\left[{\left(n-m\right)}_{mod \,N}\right]
\end{equation}
\end{small}
where $mod$ denotes modulo operation and $\ast$ is convolution symbol. Consider the DFT of $w[n]$, we have:
\begin{footnotesize} 
\begin{equation}
\begin{aligned}
W[k] &= \sum\limits_{n = 0}^{N - 1} {\sum\limits_{m = 0}^{N - 2} {h[m]x[{{(n - m)}_{mod \,N}}]} } {e^{ - j(2\pi /N)kn}}\\
&= \sum\limits_{m = 0}^{N - 1} {h[m]} {e^{ - j(2\pi /N)km}}\sum\limits_{n = 0}^{N - 1} {x[{{(n - m)}_{mod \,N}}]} {e^{ - j(2\pi /N)k(n - m)}}\\
&= H[k](\sum\limits_{n = m}^{N - 1} {x[n - m]} {e^{ - j(2\pi /N)k(n - m)}} + \sum\limits_{n = 0}^{N - 1} {x[n - m + N]} {e^{ - j(2\pi /N)k(n - m)}})\\
&= H[k](\sum\limits_{n = 0}^{N - m - 1} {x[n]} {e^{ - j(2\pi /N)kn}} + \sum\limits_{n = N - m}^{N - 1} {x[n]} {e^{ - j(2\pi /N)kn}})\\
&= H[k]\sum\limits_{n = 0}^{N - 1} {x[n]{e^{ - j(2\pi /N)kn}}} \\
&= H[k]X[k]
\end{aligned}
\end{equation}
\end{footnotesize}
where $H[k]X[k]$ is the multiplication of the two sequences in the frequency domain.

\subsection{Property of conjugate symmetric}
\noindent The conjugate symmetry property is one of the properties of the DFT. Given a signal $x[n]$, we have:
\begin{small}
\begin{equation}
X[N - k] = \sum\limits_{n = 0}^{N - 1} {x[n]{e^{ - j(2\pi /N)(N - k)n}} = } \sum\limits_{n = 0}^{N - 1} {x[n]{e^{j(2\pi /N)kn}}}  = {X^*}[k]
\end{equation}
\end{small}
In our GRLNet, we use this property to reduce learnable parameters and redundant computations.

\section{Experiments}
\label{sec: AppendixExperiments}

\subsection{Evaluation criteria}
\noindent As discussed in evaluation criteria, we use dynamic thresholds to calculate the performance of anomaly detection for the proposed GRLNet. Therefore, given a specific threshold, we can calculate the TP (True Positives), FP (False Positives), TN (True Negatives), and FN (False Negatives). Furthermore, we have:
\begin{small}
\begin{equation}
Accuracy = \frac{{TP + TN}}{{TP + TN + FP + FN}}  
\end{equation}
\end{small}
\begin{small}
\begin{equation}
Precision = \frac{{TP}}{{TP + FP}}
\end{equation}
\end{small}
\begin{small}
\begin{equation}
{\mathop{\rm Re}\nolimits} call = \frac{{TP}}{{TP + FN}}
\end{equation}
\end{small}
\begin{small}
\begin{equation}
F1-score = \frac{{2 \times Precision \times {\mathop{\rm Re}\nolimits} call}}{{Precision + {\mathop{\rm Re}\nolimits} call}}
\end{equation}
\end{small}
where the best F1-score can be obtained from the optimal global threshold. Under all possible thresholds, we can obtain a precision-recall curve (with precision as the $y$-axis and recall as the $x$-axis) and $AP = \sum\nolimits_n {(({R_n} - {P_{n - 1}})/{P_n}} )$, where ${P}_{n}$ and ${R}_{n}$ are precision and recall at the $n$-th threshold. Similarly, we can also get $TPR = TP/(TP + FN)$ and $FPR = FP/(FP + TN)$. The ROC curve with FPR as the $y$-axis and TPR as the $x$-axis. And the AUC score is the area under the ROC curve.
\begin{figure}[http]
    \centering
    \includegraphics[width=0.9\textwidth,height=45mm]{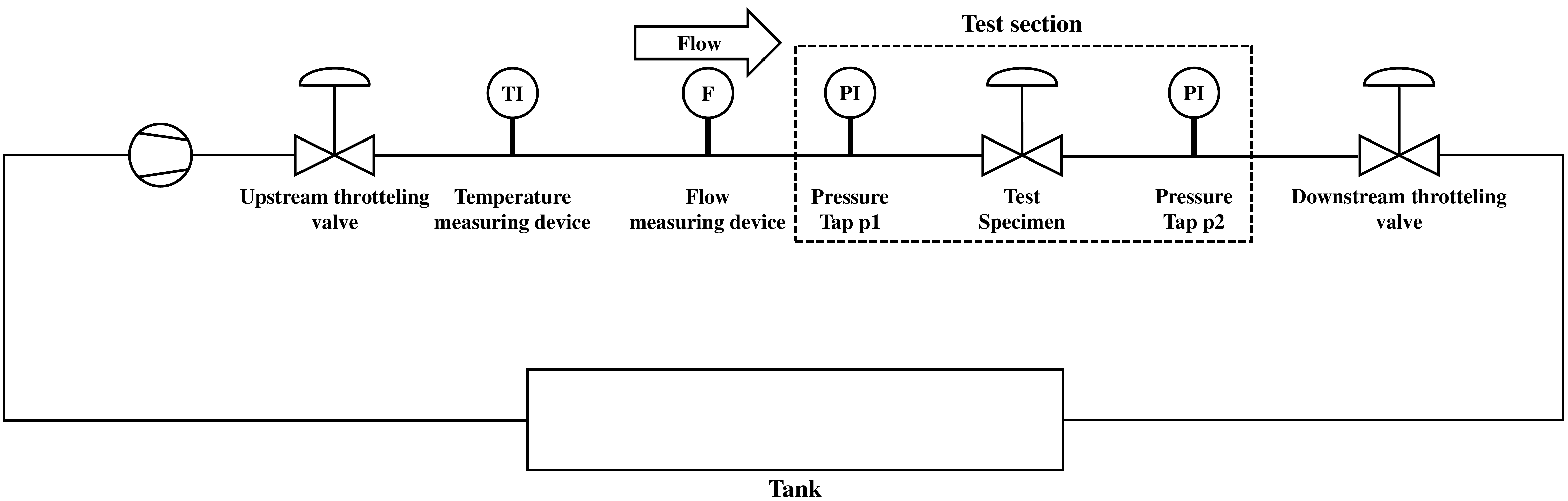}
    \caption{Schematic view of the test rack at SAMSON AG (Figure provided by SAMSON AG).}
    \label{fig: system}
\end{figure}

\subsection{Details of GRLNet}
\noindent The GRLNet consists of two generators $\mathcal{G} ^{local}$, $\mathcal{G} ^{regional}$, a discriminator $\mathcal{D}$ and a global filtering layer. $\mathcal{G} ^{local}$ and $\mathcal{G} ^{regional}$ are autoencoder structures composed of 1D convolutions and their decoder parts are both constructed by 1D transposed convolutions. The sizes of the 1D convolutional kernels of $\mathcal{G} ^{local}$ and $\mathcal{G} ^{regional}$ are $3\times 3$ and $7\times 7$, respectively. The discriminator $\mathcal{D}$ is composed of a 1D convolution and a fully connected layer. A 1D batch normalization layer and ReLU activation function are added following each of convolutional layers of $\mathcal{G} ^{local}$, $\mathcal{G} ^{regional}$ and $\mathcal{D}$. During all of these epochs, we train on Adam with learning rate set to ${10}^{-3}$ and ${10}^{-4}$ for generators ($\mathcal{G} ^{local}$ and $\mathcal{G} ^{regional}$) and discriminator $\mathcal{D}$, respectively. The adversarial loss and reconstruction loss of GRLNet are binary cross-entropy function and mean square error function, respectively. In order to better train $\mathcal{G} ^{local}$, $\mathcal{G} ^{regional}$ and $\mathcal{D}$, we propose a balanced adversarial factor $\gamma$ and a balanced discriminant factor $\alpha$. In the testing phase, we use comprehensive output factor $\beta$ to to weight the combined scores of $\mathcal{D}$ under different patterns $\mathcal{G} ^{local}$ and $\mathcal{G} ^{regional}$. The global filter layer is easily implemented in current deep learning frameworks (e.g., PyTorch, TensorFlow and so on). And the FFT can be well supported by CPU and GPU. Our source code is released at \url{https://github.com/CavitationDetection/GRLNet}.

\subsection{Datasets}
\noindent\textbf{Cavitation Datasets.} The cavitation datasets are provided by SAMSON AG in Frankfurt. The schematic of the experimental setup to collect data is shown in Figure \ref{fig: system}. And five flow status are induced in acoustic signals by varying the differential pressure at various constant upstream pressure of the control valve different operation conditions: cavitation choked flow, constant cavitation, incipient cavitation, turbulent flow and no flow (see Tables \ref{tab: CavitationDatasets-FlowStatus} and \ref{tab: CavitationDatasets-operation}). It should be noted that the Cavitation Datasets are measured by SAMSON AG in a professional environment. The training set consists of non-cavitation, and the test set contains three types of cavitation and non-cavitation. For detailed dataset statistics shown in Table \ref{tab: training and test sets}.
\input{tab/Appendix/FlowSatus}

\input{tab/Appendix/Operations}
\input{tab/Appendix/cavitation_training_test}

\noindent\textbf{MIMII.} In our research, the dataset is split into a training set and a test set. The test set is consisted of all abnormal acoustic signals and the same number of normal acoustic signals. The other normal acoustic signals are regarded as the training set (see Table \ref{tab: MIMII training and test sets}).
\input{tab/Appendix/MIMII_training_test}

\subsection{Ablation analysis}
\noindent\textbf{Analysis of different intensity noise.} We compare the anomaly detection capability of GRLNet with different intensity noise in cavitation2017 (anomaly percentage of 50$\%$), see Table \ref{tab: results different SNR}.
\input{tab/Appendix/results_different_SNR}

%% file: tab/Appendix/FlowSatus.tex
\begin{table}[htbp]
\centering
\caption{Cavitation datasets content details.}
\label{tab: CavitationDatasets-FlowStatus}
\small
\setlength{\tabcolsep}{0.3mm}{
\begin{tabular}{lccccc}
\toprule 
\multirow{2}{*}{Dataset} & \multicolumn{3}{c}{Abnormal} & \multicolumn{2}{c}{Normal}  \\ 
\cmidrule{2-6} 
& \multicolumn{1}{l}{choked flow} & \multicolumn{1}{l}{constant} & \multicolumn{1}{l}{incipient} & \multicolumn{1}{l}{turbulent} & \multicolumn{1}{l}{no flow} \\ 
\midrule
Cavitation2017           & 72    & 93   & 40   & 118   & 33  \\ 
Cavitation2018           & 148   & 396  & 64   & 183   & 15  \\ 
Cavitation2018-noise     & 40    & 40   & 40   & 40    & 0  \\ 
\bottomrule
\end{tabular}}
\end{table}

%% file: tab/Appendix/Operations.tex
\begin{table}[htbp]
\centering
\caption{Details of cavitation datasets valve operation.}
\label{tab: CavitationDatasets-operation}
\small
\setlength{\tabcolsep}{0mm}{
\begin{tabular}{lcc}
\toprule 
\multirow{2}{*}{Dataset} & \multicolumn{2}{c}{Operation}\\ 
\cmidrule{2-3} 
& Valve stroke (mm) & \multicolumn{1}{l}{Upstream pressure (bar)} \\ 
\midrule
Cavitation2017        & {[}15,13.5,11.25,7.5,3.75,1.5,0.75{]} & {[}10,9,6,4{]}   \\
Cavitation2018        & {[}60,55,45,30,25,15,6{]}             & {[}10,6,4{]}     \\
Cavitation2018-noise  & 15                                    & 10              \\ 
\bottomrule
\end{tabular}
}
\end{table}

%% file: tab/Appendix/cavitation_training_test.tex
\begin{table}[htbp]
\centering
\caption{Details of the training and test sets. $(\cdot )$ denotes the number after the sliding window (window size is 466944).}
\label{tab: training and test sets}
\footnotesize
\setlength{\tabcolsep}{0.05mm}{
\begin{tabular}{llcccccc}
\toprule 
\multicolumn{1}{c}{\multirow{3}{*}{Dataset}} & \multirow{3}{*}{Train}  & \multicolumn{6}{c}{Test}  \\ 
\cmidrule{3-8} 
\multicolumn{1}{c}{}&& \multicolumn{1}{l}{\multirow{2}{*}{Non-cavitation}} & \multicolumn{5}{c}{Cavitation (\%)}\\ 
\cmidrule{4-8} 
\multicolumn{1}{c}{}&  & \multicolumn{1}{l}{} & \multicolumn{1}{l}{10\%} & \multicolumn{1}{l}{20\%} & \multicolumn{1}{l}{30\%} & \multicolumn{1}{l}{40\%} & \multicolumn{1}{l}{50\%} \\ 
\midrule
Cavitation2017 & \multicolumn{1}{c}{121 (1210)} & 30 (300)  & 3 (30)  & 6 (60) & 12 (120)  & 20 (200)  & 30 (300)           \\
Cavitation2018 & \multicolumn{1}{c}{158 (13114)} & 40 (3320)  & 4 (332)  & 10 (830) & 17 (1411)  & 26 (2158)  & 40 (3320)   \\
Cavitation2018-noise& \multicolumn{1}{c}{32 (2656)}  & 8 (664)  & 1 (83) & 2 (166)  & 3 (249)  & 5 (415)  & 8 (664)         \\ 
\bottomrule
\end{tabular}
}
\end{table}

%% file: tab/Appendix/MIMII_training_test.tex
\begin{table}[htbp]
\centering
\caption{Details of the training and test sets on MIMII.}
\label{tab: MIMII training and test sets}
\footnotesize
\setlength{\tabcolsep}{0.8mm}{
\begin{tabular}{cccc}
\toprule 
\multicolumn{2}{l}{Machine type / model ID} & Training set & Test set \\ 
\midrule
\multirow{4}{*}{Valve}            & 00      & 872          & 238      \\
                                  & 02      & 588          & 240      \\
                                  & 04      & 880          & 240      \\
                                  & 06      & 872          & 240      \\ 
\midrule
\multirow{4}{*}{Pump}             & 00      & 863          & 286      \\
                                  & 02      & 894          & 222      \\
                                  & 04      & 602          & 200      \\
                                  & 06      & 934          & 204      \\ 
\midrule
\multirow{4}{*}{Fan}              & 00      & 604          & 814      \\
                                  & 02      & 657          & 718      \\
                                  & 04      & 685          & 696      \\
                                  & 06      & 654          & 722      \\ 
\midrule
\multirow{4}{*}{Slide rail}       & 00      & 712          & 712      \\
                                  & 02      & 801          & 534      \\
                                  & 04      & 356          & 256      \\
                                  & 06      & 445          & 178      \\ 
\bottomrule
\end{tabular}
}
\end{table}

%% file: tab/Appendix/results_different_SNR.tex
\begin{table}[htbp]
\centering
\caption{AUC and ${F}_{1}$ score results for different intensity noise.}
\label{tab: results different SNR}
\footnotesize
\footnotesize
\setlength{\tabcolsep}{1mm}{
\begin{tabular}{cccccccc}
\toprule 
\multirow{2}{*}{Results} & \multicolumn{7}{c}{Different intensity noise (dB)} \\ 
\cmidrule{2-8} 
                         & -6   & -4   & -2   & 0  & 2  & 4  & 6  \\ 
\midrule 
AUC                      & 0.9753  & 0.9747  & 0.9756  & 0.9750   & 0.9756   & 0.9657  & 0.9748    \\
F1 score                 & 0.9831  & 0.9699  & 0.9831  & 0.9633   & 0.9831   & 0.9616  & 0.9699   \\ 
\bottomrule
\end{tabular}
}
\end{table}